\documentclass[aps,prl,twocolumn,showkeys,groupedaddress,nofootinbib]{revtex4}
\usepackage[pdftex]{graphicx}

\begin{document}
\title{Integration of graphene-based superconducting quantum circuits in 3D cavity}

\author{Kuei-Lin Chiu$^{1,*}$, Youyi Chang$^{1}$, Avishma J. Lasrado$^{1}$, Cheng-Han Lo$^{1}$, Yung-Hsiang Chen$^{1}$, Tao-Yi Hsu$^{1}$, Yen-Chih Chen$^{1}$, Yi-Chen Tsai$^{3}$, Samina$^{2}$, Yen-Hsiang Lin$^{2}$, Chung-Ting Ke$^{3,*}$}

\affiliation{$^1$Department of Physics, National Sun Yat-Sen University, Kaohsiung 80424, Taiwan}
\affiliation{$^2$Department of physics, National Tsing-Hua University, Hsinchu 300044, Taiwan}
\affiliation{$^3$Institute of Physics, Academia Sinica, Taipei 115201, Taiwan}

\date{\today}

\begin{abstract}
Integrating 2D materials into circuit quantum electrodynamics (c-QED) devices is an emerging filed in recent years. This integration not only facilitates the exploration of potential applications in quantum information processing but also enables the study of material's fundamental properties using microwave techniques. While most studies employ 2D coplanar architectures with scalability potential, 3D cavity based c-QED devices, due to their simpler design, offer the advantage of a quicker turnaround to probe the composite Josephson junctions (JJs). Here, we construct the first flux-tunable, 3D cavity-compatible superconducting quantum circuit based on 2D materials, featuring a graphene superconducting quantum interference device (SQUID) shunted by a capacitor that is accessible by both DC and microwave probes. We have shown how flux-modulated cavity frequency can be linked to the SQUID critical current under the influence of Fraunhofer pattern. In addition, we extracted the symmetry information of the SQUIDs based on DC analysis, and correlated this with the flux-modulated cavity frequency observed in microwave measurements. Our platform can extend to topological materials, holding the prospect of establishing valid topological JJs with DC probe while allowing fast microwave probe to avoid quasiparticle poisoning.

\end{abstract}

\maketitle

\def\thefootnote{*}\footnotetext{Corresponding author: K. L. Chiu (klc@mail.nsysu.edu.tw)}\def\thefootnote{\arabic{footnote}}
\def\thefootnote{*}\footnotetext{Corresponding author: C. T. Ke (ctke@gate.sinica.edu.tw)}\def\thefootnote{\arabic{footnote}}

\section{I. Introduction}

Quantum materials, owing to their rich internal degrees of freedom unexplored, provide versatile functions to be integrated as a main ingredient in quantum information devices \cite{Kroll2018,Wang2019,LucasCasparis2018,AlbertHertel2022,Lo2023,Larsen2015,Lange2015,Huo2023,Chiu2017,Liu2019,Chiu2020b,Siddiqi2021}. Graphene, two-dimensional electron gas (2DEG) and 1D nanowires, due to their low dimension hence gate-tunable nature, can serve as normal metals in S-N-S junctions in gatemons (S denotes a superconductor and N denotes a normal metal) \cite{Kroll2018,Wang2019,LucasCasparis2018,AlbertHertel2022,Lo2023,Larsen2015,Lange2015,Huo2023,Xia2024,Sagi2024,Zhuo2023,Mergenthaler2021}. Similarly, S-I-S junctions (I denotes an insulator) using multi-layer semiconducting MoS$_2$ as a tunnel barrier have been integrated into qubits \cite{Lee2019}. Furthermore, using topological materials as the weak link in JJs may provide a different route to realizing the exotic Majorana bound states (MBS) where 4$\pi$-period phase modulation is expected \cite{Fu2009}. Relevant studies such as integration of Weyl semimetal MoTe$_2$ and topological insulator (Bi$_{0.06}$Sb$_{0.94})_2$Te$_3$ in superconducting quantum circuits have been reported \cite{Chiu2020,Schmitt2022}. On the other hand, 2D materials can be directly integrated with a resonator, forming a quantum harmonic oscillator instead of the quantum anharmonic oscillator in transmon cases. One example is the use of an embedded graphene JJ in coplanar superconducting resonators, which allows for simultaneous in-situ DC and RF measurements to study the junction inductance \cite{Schmidt2018,Schmidt2020a,Tanaka2024}. In addition, in inductively coupled resonator-type devices, it has been shown that microwave measurements can be used to probe the Andreev bound states (ABSs) in the JJs \cite{Hinderling2023,Hays2018}. In a particular layout, it is also possible to perform in-situ qubit measurements while inferring the transport properties of the compositing JJ in a transmon device \cite{Kringhoj2020}.

While the exploration of their potential for quantum information processing is intriguing, 2D material-based c-QED devices offer the additional advantage of using microwaves to probe JJ's fundamental properties that are inaccessible through DC transport measurements. In topological junctions possessing exotic 4$\pi$ periodic supercurrent, transport measurements can reveal the existence of MBS through the missing n=1 Shapiro steps \cite{Wiedenmann2016,Bocquillon2017,Li2018a,Wang2018a}. However, due to the slow probing speed, they often suffer from quasiparticle poisoning, which tend to restore the unconventional 4$\pi$ periodic supercurrent to trivial 2$\pi$ periodicity \cite{Badiane2013,Sun2022}. In contrast, microwave techniques typically operate on a microsecond time scale \cite{Krantz2020}, which is fast enough to probe dynamic processes such as quasiparticle tunneling that often occur on time scales between $\mu$s and ms \cite{Rainis2012a,Karzig2021,Sun2022}. Theory has predicted that transmons consisting of topological SQUIDs exhibit a 4$\pi$ periodic flux modulation in qubit transition frequency \cite{Sun2022}, enabling time-domain spectroscopy to manipulate qubit states and probe MBS-related physics faster than the quasiparticle poisoning rate \cite{Sun2022}. Inspired by this, it is desirable to construct SQUID-based c-QED devices that can be accessed by Shapiro step measurements (DC transport) while enabling microwave techniques to probe their flux-tunability.

In this paper, we construct 3D cavity-compatible c-QED devices consisting of graphene SQUIDs as prototypes to explore the aforementioned concept. While most studies were based on 2D coplanar waveguides \cite{Kroll2018,Wang2019,LucasCasparis2018,AlbertHertel2022,Lo2023,Larsen2015,Lange2015,Huo2023,Schmidt2018,Schmidt2020a,Tanaka2024}, we employ 3D cavities as they possess several advantages over their 2D counterparts. 3D cavities provide a simpler electromagnetic environment for superconducting qubits, thus greatly reducing dielectric loss and improving qubit performance \cite{Paik2011}. In addition, the adjustment of c-QED parameters, such as cavity decay rate $\kappa$ and qubit-cavity coupling strength $g$, can be handily implemented by altering the readout pin length and position of the qubit within the cavity \cite{Zoepfl2017}, which is not achievable with a 2D readout resonator once its pattern has been fabricated. Most importantly, 3D cavity devices are relatively simple in design and fabrication, without concerning the quality of 2D resonators in each fabrication round, which significantly improves the turnover rate of various measurements. 

Our graphene-based superconducting quantum circuit consists of a pair of capacitor pads, which serve as the antenna to interact with microwave in a 3D copper cavity as well as for the bond pads in DC transport measurements. We find correlations between DC and RF measurements, including the Fraunhofer pattern, flux-modulated supercurrent, and the corresponding resonator frequency, across three devices with varying degrees of SQUID symmetry. We demonstrate the first 2D material-based flux-tunable c-QED device integrated into 3D cavities, which can be extended to topological materials. This platform holds the potential to establish valid topological JJs $via$ Shapiro step measurements while performing time-domain microwave measurements for dynamic studies.

\begin{figure}[!t]	
\includegraphics[scale=0.45]{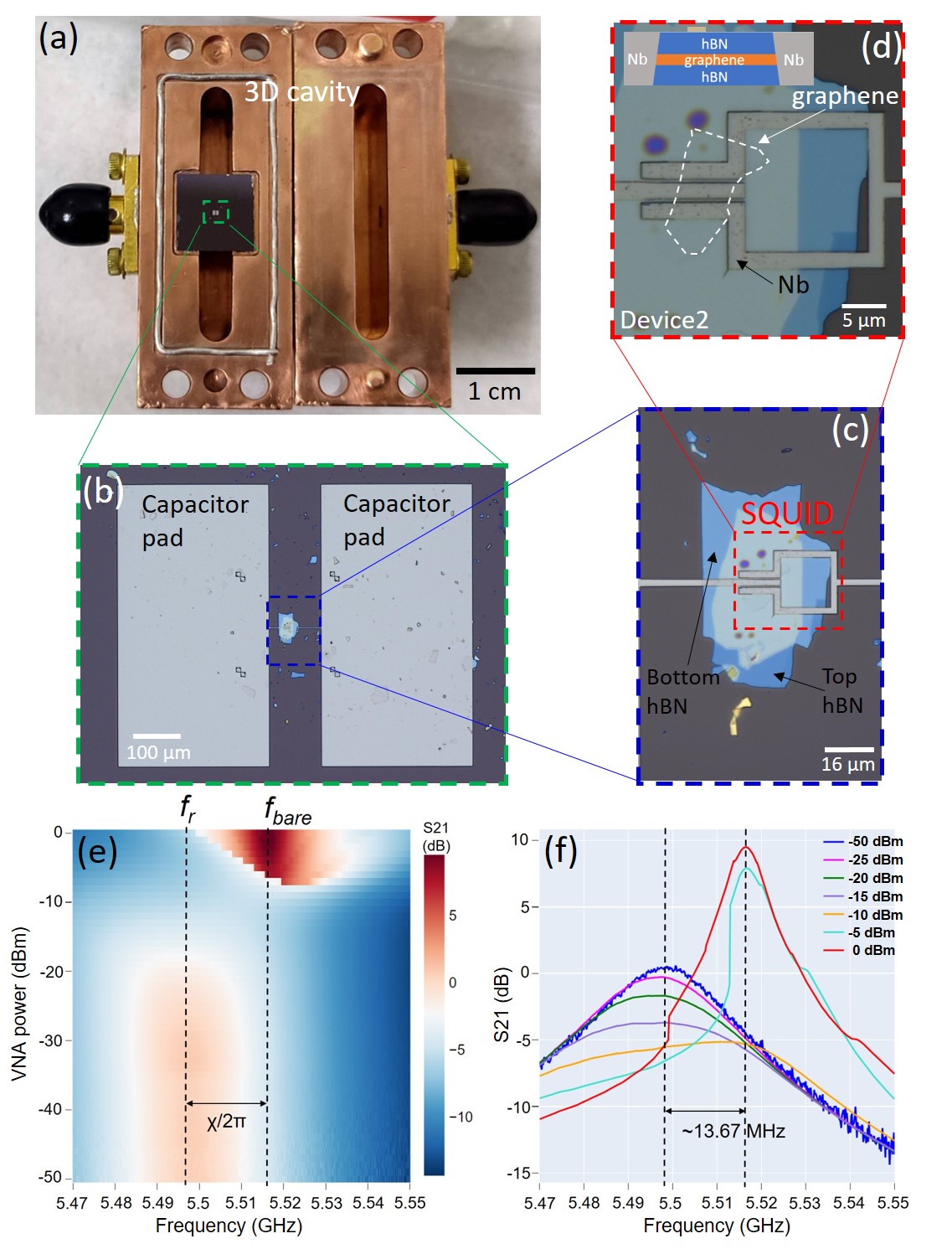}
\caption{Optical micrograph and power dependence measurements of a flux-tunable graphene-based superconducting quantum circuit coupled to a 3D copper cavity. (a) 3D copper cavity with a device residing in the center. (b) Optical micrograph of the device, consisting of the shunting capacitor with SQUID loop in the center. (c) Optical micrograph showing the SQUID made of graphene and superconductor Nb. The graphene flake was sandwiched between two hBN layers [inset in (d)] before making edge contacts of Nb. (d) The zoom-in image of (c) showing the geometry of the SQUID, which is formed by a square loop (line width: 2 $\mu$m; inner area: 16 $\mu$m $\times$ 16 $\mu$m) with encapsulated graphene linking two 500 nm gaps. The inset shows the hBN/graphene/hBN sandwich structure with edge contacts. The device shown in (a) - (d) is referred as device 2 in this work. (e) Power dependence measurement of device 1 [Fig. \ref{FigS2} (c)], confirming the presence of graphene JJs by showing a dispersive shift $\chi/2\pi$. (f) Linecuts in (e) at different powers.}      
\label{Fig1}
\end{figure}

\section{II. Device fabrication and measurement setup}
Fig. \ref{Fig1} shows the optical micrograph of our flux tunable graphene-based superconducting quantum circuits coupled to a 3D copper cavity. This 3D cavity-compatible quantum circuit consists of a SQUID made of two Nb-Graphene-Nb junctions, with an enclosed loop of Nb at a nominal size of 16 $\times$ 16 $\mu m^2$ [Fig. \ref{Fig1} (d)]. To preserve the graphene quality, the exfoliated graphene is encapsulated by hexagonal Boron Nitride (hBN) [Fig. \ref{Fig1} (c)], and the SQUID is fabricated based on the edge contact techniques \cite{Wang2013}, as described in Appendix A. The SQUID is shunted by a capacitor formed between two rectangular capacitor pads as shown in Fig. \ref{Fig1} (b), which is in analogy to transmon architecture. In microwave measurements, these two capacitor pads act as an antenna to interact with electromagnetic field in 3D cavities; while in DC transport measurements, these two capacitor pads are used as bond pads.

The measurement scheme for microwave and DC transport exploited in this work is illustrated in Fig. \ref{FigS3} (a) and (b), respectively. For microwave measurements, as shown in Fig. \ref{Fig1} (a), the devices were placed in a two-ports 3D copper cavity to provide good thermal conductivity and to allow externally applied magnetic fields to thread through the cavity for flux-tuning. For design and calibration of 3D cavities, performing transmission (S$_{21}$) and DC transport measurements, please refer to Appendix B. In this work, three devices labeled as 1, 2 and 3 were characterized, and they were placed in a 5.5 GHz (device 1 and 2) cavity and a 6.03 GHz (device 3) cavity for microwave measurements, respectively. In addition, for all three devices, we first performed microwave measurements, followed by DC transport measurements in different cool-downs, as detailed in the following sections.

\section{III. Device characterization and results}
We first performed the power dependence measurements on device 1, in which S$_{21}$ measured through the two-ports 3D cavity was recorded as a function of readout power and frequency, as shown in Fig. \ref{Fig1}(e). This is a conventional way to confirm the existence of JJ \cite{Reed2010}, where the resonant frequency of the cavity ($f_{r}$) shifts toward the qubit frequency at large enough power. Here, we note that we have not demonstrated temporal coherence or energy-level spectroscopy in our devices, which does not justify the devices as qubits. However, for simplicity we still use terms, such as qubit frequency or qubit-cavity coupling, in the following content. Fig. \ref{Fig1}(f) shows the linecuts at different readout powers. At lower powers from -50 dBm to -15 dBm, the cavity frequency centers around 5.5 GHz with an asymmetric line shape, indicating a canonical behavior of a Kerr-Duffing oscillator \cite{Reed2014a}. After undergoing a broad range (-15 dBm $\leq$ P $\leq$ -8 dBm), the cavity response re-appears with a narrower line shape of Lorentzian and a frequency shifted to 5.516 GHz at P = 0 dBm, which is known as the bare cavity frequency ($f_{bare}$). The dispersive shift $\chi/2\pi$ of around 13.67 MHz between cavity's frequency at low power ($f_r$) and at high-power ($f_{bare}$) allows us to estimate the qubit-cavity coupling strength $g$ $via$ $\chi = g^2/\Delta$ and $\Delta = 2\pi(f_{r} - f_q)$, where $f_q$ is qubit frequency. To deduce $f_q$, we have performed two-tone measurements (data not shown), but no qubit transition was found in the probed frequency range of 4 to 12 GHz. Although we cannot obtain direct information about $f_q$ $via$ two-tone, we have relied on critical current in transport measurements to estimate both $f_q$ and $g$, as discussed in Appendix D. 

\begin{figure*}[!t]	
\includegraphics[scale=0.6]{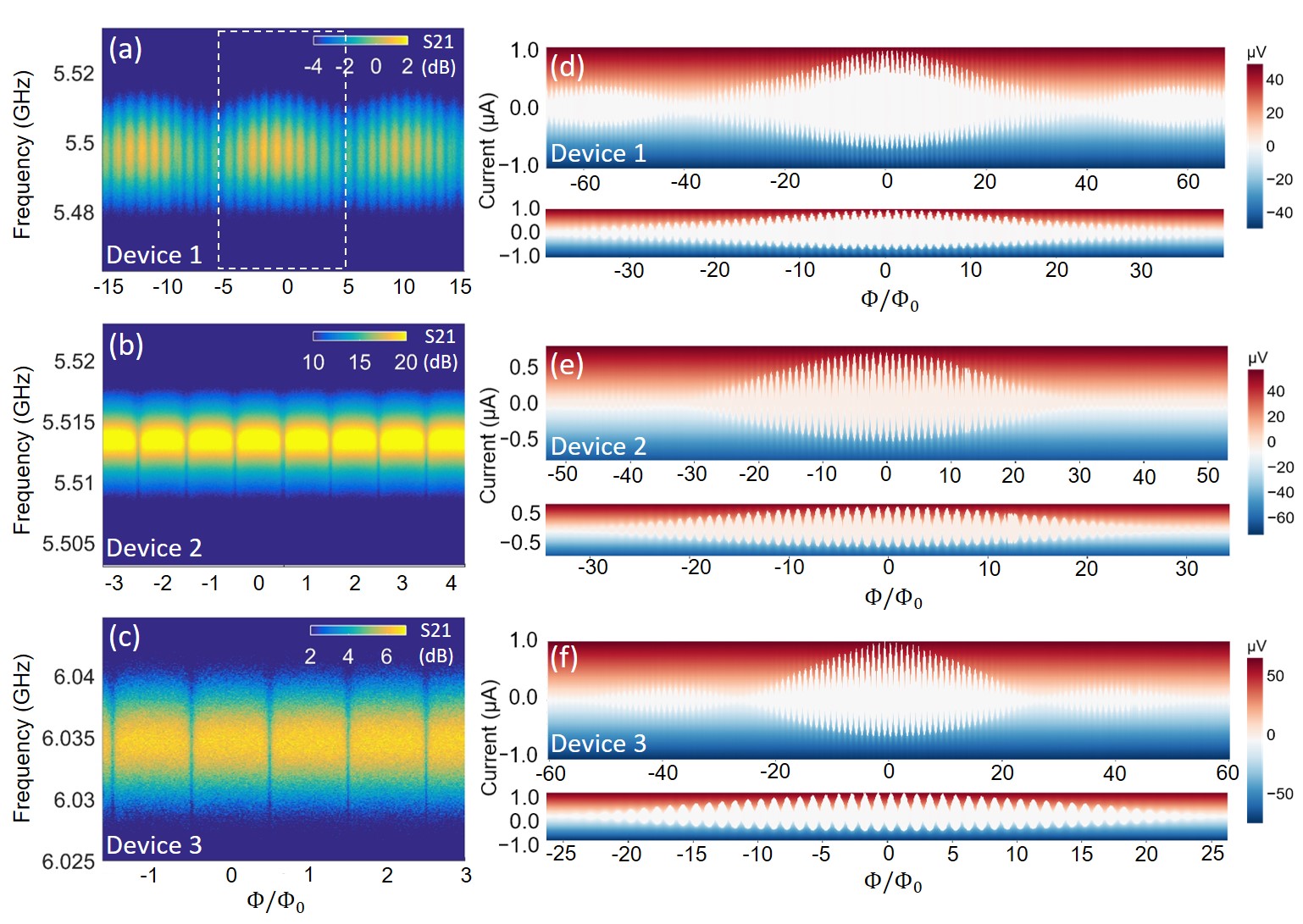}
\caption{Flux modulation of cavity frequency (microwave measurements) and critical current (DC transport measurements) of our graphene-based superconducting quantum circuits in different cool-down. (a) - (c) Flux modulation of cavity frequency for device 1, 2 and 3. The dashed rectangle in (a) indicates 10 SQUID oscillations in a Fraunhofer lobe. (d) - (f) Flux modulation of SQUID critical current for device 1, 2 and 3. The colorscale represents the DC voltage across the junctions. The bottom panels show the SQUID oscillations in the Fraunhofer central lobe for each device.}
\label{Fig3}
\end{figure*} 

\begin{figure*}[!t]	
\includegraphics[scale=0.52]{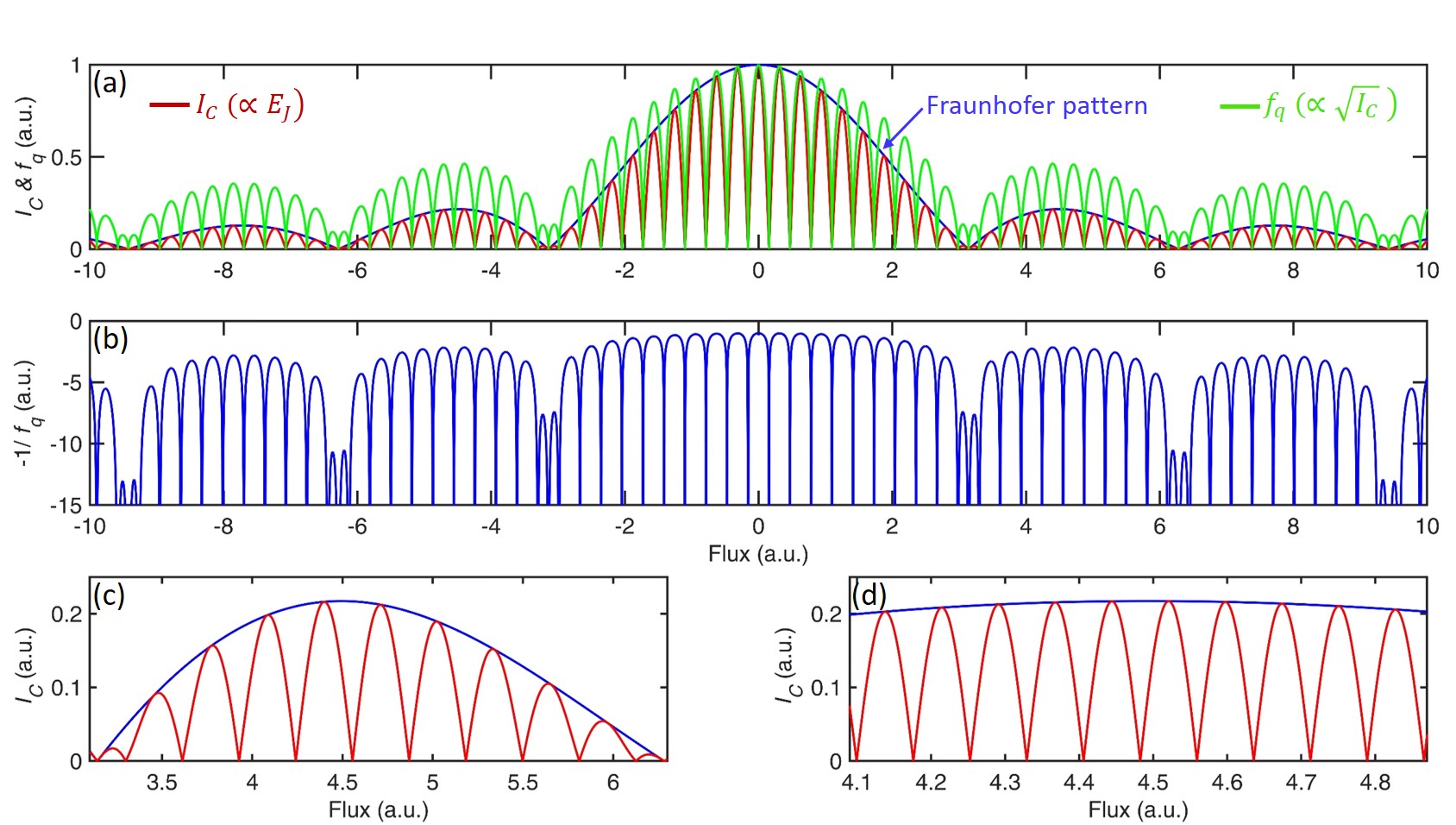}
\caption{Schematic simulation of SQUID critical current, qubit frequency and dispersive shift modulated by flux under the influence of Fraunhofer effect. More details can be found in Appendix C. (a) SQUID critical current ($I_c$, red) and qubit frequency ($f_q$, green) are periodically modulated by flux, with the blue curves indicating the Fraunhofer pattern. Note that we have set 10 oscillations in a sub-lobe (20 oscillations in the central lobe) to emulate the situation shown in Fig. \ref{Fig3}(a). (b) Dispersive shift modulated periodically by flux. (c) Showing 10 oscillations in critical current when a Fraunhofer sub-lobe contains 10 SQUID oscillations. (d) Showing 10 oscillations in critical current when a Fraunhofer sub-lobe contains 40 SQUID oscillations. (c) and (d) are zoom-in images of Fig. \ref{FigS5}(a) and (b) with a selected range.}
\label{Fig4}
\end{figure*}

Microwave measurements of flux-tuning was executed by measuring the resonant cavity frequency at a low readout power (-55 dBm) while tuning the DC source current passing through the home-made superconducting coil around the copper 3D cavity. The flux-modulated resonant cavity frequency for device 1, 2 and 3 is shown in Fig. \ref{Fig3}(a), (b) and (c), respectively. Interestingly, while a periodic modulation of cavity frequency was observed in device 2 and 3 [Fig. \ref{Fig3}(b) and (c)], as being commonly seen in conventional Al$/$Al$_2$O$_3$-based transmons \cite{Chow2010}, there is an additional larger-period flux modulation observed in device 1, as indicated by the dashed box in Fig. \ref{Fig3}(a). This larger-period modulation is superimposed on a fine modulation presumably from SQUID critical current oscillation, with a period ratio of 10 to 1. We attribute this additional modulation to the Fraunhofer oscillation resulted from the composing graphene JJs (see detailed discussions in Appendix C). The Fraunhofer pattern-modulated SQUID oscillations (see equation (1)) are illustrated in Fig. \ref{Fig4}(a), with the SQUID loop area set to be 10 times the junction area, to emulate the situation observed in Fig. \ref{Fig3}(a). As shown in Fig. \ref{Fig4}(a), the Fraunhofer pattern indicated by the blue solid lines is modulating the SQUID critical current oscillations represented by the red solid lines. We can relate the Fraunhofer effect-mediated SQUID critical current to various physical parameters. Since Josephson energy $E_J$ = $\frac{\Phi_0I_C}{2\pi}$ is proportional to $I_C$ while qubit frequency $f_q$ $\approx$ $\sqrt{8 E_J E_C}/h$ is proportional to $\sqrt{I_C}$, their flux modulation under Fraunhofer effect is shown in red and green solid lines in Fig. \ref{Fig4}(a), respectively. By controlling the magnetic flux threading through the SQUID loop, we are able to modify the critical current, hence $E_J$ and $f_q$, which leads to different dispersive shift $\chi$ to be measured in cavity frequency. In Fig. \ref{Fig4}(b), we present the schematically simulated dispersive shift under the influence of Fraunhofer effect (see discussions in Appendix C). Due to the effects of square root and reciprocal transformations, the large variations in the height of Fraunhofer lobes [blues curve in Fig. \ref{Fig4}(a)] have become a relatively small variations in dispersive shift as shown in Fig. \ref{Fig4}(b), which agrees with the observations in Fig. \ref{Fig3}(a). 

To gain more insight into how SQUID critical current correlates with the microwave measurement, we conducted DC transport measurements on device 1, 2 and 3, as shown in Fig. \ref{Fig3}(d), (e) and (f), respectively. All devices exhibited SQUID modulations with a Fraunhofer pattern, showing 82, 61, and 55 oscillations in the central lobe for devices 1, 2, and 3, respectively [bottom panels of Fig. \ref{Fig3}(d), (e) and (f)]. Notably, for device 1, Fig. \ref{Fig3}(d) indicates that the number of SQUID oscillations in the Fraunhofer sub-lobe is 41, which is about four times the number observed in the microwave measurement shown in Fig. \ref{Fig3}(a). In later analysis (see Appendix E), we found that SQUID's JJs in device 1 are asymmetric with a ratio $I_{C1}$ : $I_{C2}$ = 1:4, which coincides with the ratio 10:41 comparing microwave with DC results. Thus, we attribute the Fraunhofer modulation observed in microwave (DC) measurements to the larger (smaller) junction in the asymmetric SQUID, with an area ratio of 1:10 (1:40) relative to the SQUID loop, as indicated by JJ$_{2(1)}$ in Fig. \ref{FigS5}(c). It should be noted that there may exist a large magnetic field offset between DC and microwave measurements. While B-filed is applied through a superconducting magnet in DC setup, in microwave setup it is generated by a homemade superconducting coil, which is more prone to trapping environmental magnetic flux. Consequently, this could result in probing different regimes of the Fraunhofer pattern in each setup with a very different offset value, giving rise to the modulation of the Fraunhofer sub-lobe, rather than the central lobe, observed in the microwave measurements, as shown in Fig. \ref{Fig3}(a). The Fraunhofer pattern was not pronounced in the microwave measurements of device 2 and device 3 [Fig. \ref{Fig3}(b) and (c)]. In contrast, the transport data presented in Fig. \ref{Fig3}(e) and Fig. \ref{Fig3}(f) show clear Fraunhofer modulation, with the central lobes containing a large number of SQUID oscillations. The lack of Fraunhofer modulation in microwave measurements for device 2 and 3 can be attributed to their more symmetric SQUIDs (see Appendix E) whose JJs are both small, so there is no Fraunhofer modulation from a much larger JJ as in the case of device 1. An example is illustrated in Fig. \ref{FigS5} (a) and (b), where different area ratios between the junction and the SQUID loop result in 10 and 40 SQUID oscillations within a Fraunhofer sub-lobe. Thus, when comparing 10 SQUID oscillations in each case, the Fraunhofer modulation is less pronounced for a larger SQUID loop-to-junction area ratio, as illustrated in Fig. \ref{Fig4}(c) and (d), respectively.



\begin{figure*}[!t]	
\includegraphics[scale=0.57]{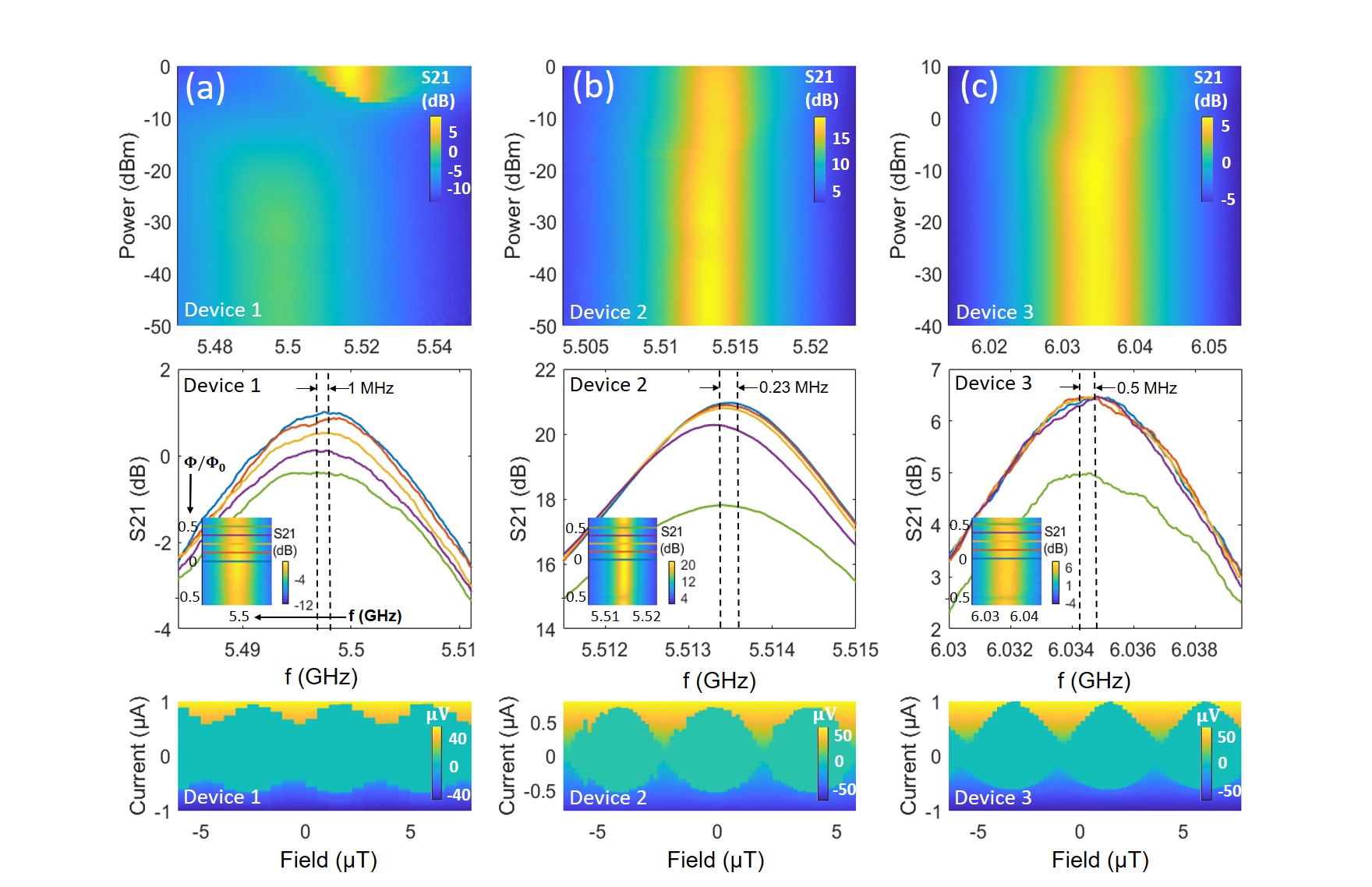}
\caption{Comparison of power dependence, flux tunability, and SQUID critical current oscillation across three devices. (a) Top panel: Power dependence measurements for device 1. Middle panel: Linecut along the frequency direction indicated by the colored lines in the inset. Inset: Flux modulation of cavity frequency across one period. The linecuts are equally spaced within half a flux period from 0 to 0.5 $\Phi_0$. Bottom panel: Oscillation of SQUID critical current for device 1 in a small B-field range. The colorscale represents the DC voltage across the junctions. (b) The same as (a) but for device 2. (c) The same as (a) but for device 3.}
\label{Fig5}
\end{figure*}

\begin{figure*}[!t]	
\includegraphics[scale=0.47]{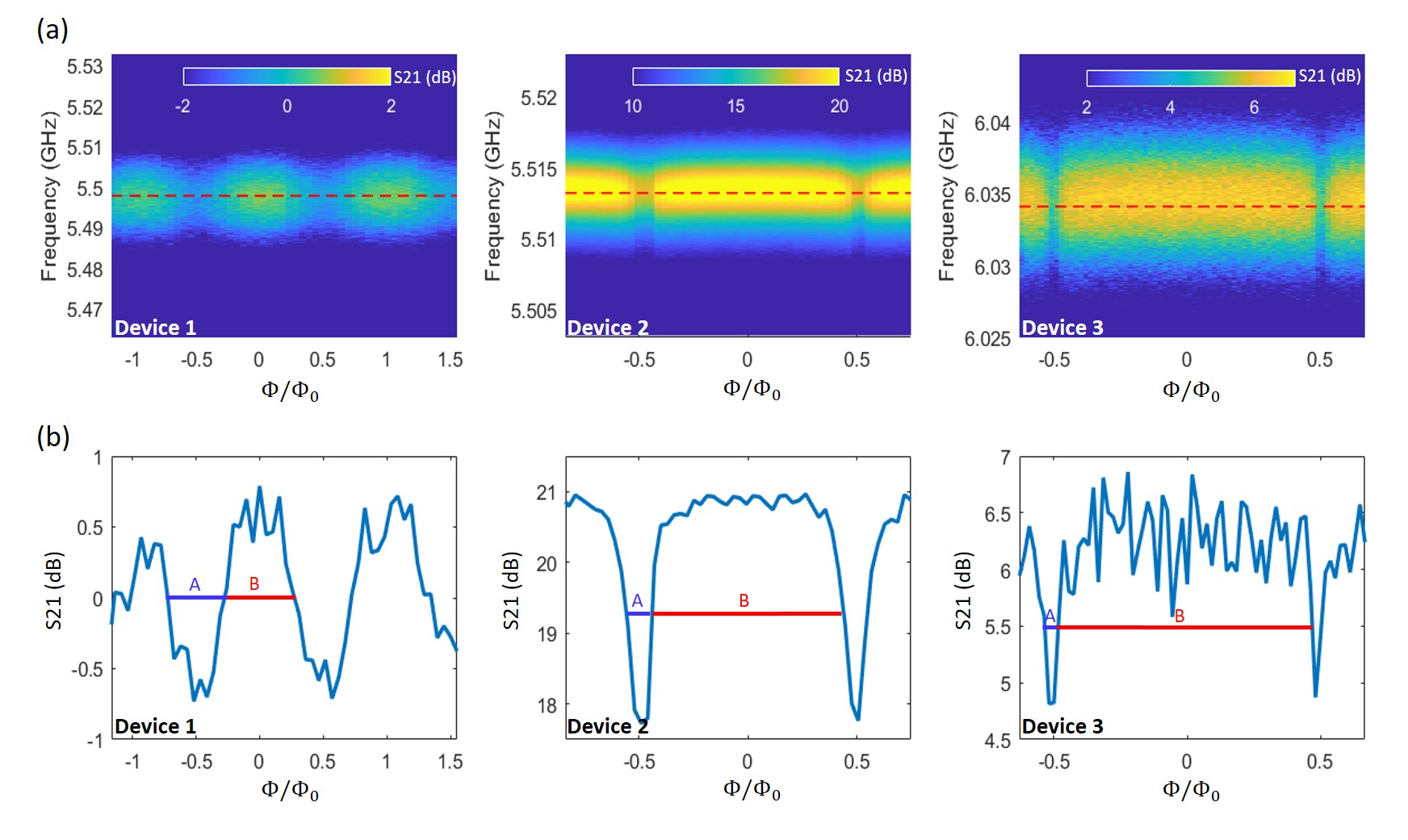}
\caption{(a) Flux modulation of cavity frequency for device 1, 2 and 3. (b) The corresponding linecuts along the red dashed lines in (a). For device 2 and 3, the lines labeled by A denote the flux range where $f_{q}$ $<$  $f_{r}$, while the lines labeled by B denote the flux range where $f_{q}$ $>$  $f_{r}$. For device 1, A and B denote the flux ranges corresponding to the downward and upward behavior of the oscillating qubit frequency, respectively.}      
\label{Fig6}
\end{figure*}

We further compare the power dependence, flux-tunability of cavity frequency and DC critical current of SQUID for three devices, as shown in Fig. \ref{Fig5}. In the top panels of Fig. \ref{Fig5}, the dispersive shift is 13.67 MHz, 0.48 MHz and 0.68 MHz for device 1, 2 and 3; while in the middle panels the maximal flux-tunability of cavity frequency is 1 MHz, 0.23 MHz and 0.5 MHz for device 1, 2 and 3, respectively. In Appendix D, we have estimated the qubit-cavity coupling strength $g$ based on the SQUID critical current extracted from the sub-lobe of Fraunhofer pattern. The estimated coupling strength $g/2\pi$ are 318.9 MHz, 41.76 MHz, and 68.5 for device 1, 2, and 3, respectively. The exact cause of the variation in $g$ is not entirely clear. We speculate that the remaining metallic graphite pieces interact with microwave, thus amend the distribution of electromagnetic field inside the cavity and enhance the local electric field, causing the variation in qubit-cavity coupling strength observed across different devices. The estimated $g$ for device 2 and 3 is small, which can account for the relatively small fluxtunability. However, the estimated $g$ for device 1 is relatively large, yet the flux-tunability remains limited. In addition, if the flux-modulated qubit frequency intersects with cavity frequency, a large Rabi splitting is expected, but this was not observed in Fig. \ref{Fig3} (a). We attribute this discrepancy to the asymmetry of the SQUID in device 1 compared to device 2 and device 3, as revealed in the bottom panels of Fig. \ref{Fig5}, which display three SQUID oscillations around zero field for each device. The asymmetry of the SQUID and the resulting skewed current phase relation (CPR) have been studied in graphene and InAs junctions before \cite{Thompson2017,Mayer2020}. It has been shown that the minimum value of SQUID critical current $I_{C,min}$ depends solely on the asymmetry of JJs \cite{Clarke2004}. With increasing asymmetry between junctions, $I_{C,min}$ increases, leading to a reduction of modulation depth of SQUID critical current ($I_{C,max}$$-$$I_{C,min}$)/$I_0$, where $I_0$ is the average critical current of the two junctions \cite{Clarke2004}. This situation can be clearly observed in the bottom panels of Fig. \ref{Fig5}, where $I_{C,min}$ around 0.6 $\mu$A for device 1 is higher compared to 0.1 $\mu$A and 0.2 $\mu$A for device 2 and 3. This results in a smaller modulation depth of the SQUID critical current in device 1 compared to devices 2 and 3, indicating a highly asymmetric SQUID in device 1 relative to the others (see relevant discussions in Appendix E). Transmons consisting of asymmetric SQUIDs have been investigated before in ref. \cite{Krantz2020}. The modulation of $E_J$ with applied flux $\Phi$ in such a transmon is described by $E_J (\Phi)$ = $E_{J\Sigma}$ $\sqrt{cos^2(\Phi)+d^2sin^2(\Phi)}$, where $E_{J\Sigma}$ = $E_{J1}$ + $E_{J2}$ and $d$ = ($\gamma$ $-$ 1)/($\gamma$ + 1) is the junction asymmetry parameter, with $\gamma$ = $E_{J2}$/$E_{J1}$ (subscript 1 and 2 denote the JJ index) \cite{Krantz2020}. A large junction asymmetry parameter will result in qubit frequency oscillating at its maximal frequency with suppressed flux modulation depth \cite{Krantz2020}. As shown in Fig. \ref{FigS7} (a), the qubit frequency for device 1 oscillating between 12.935 GHz and 10.02 GHz is much higher than the cavity frequency around 5.5 GHz, leading to the small flux-tunability in cavity frequency despite a large qubit-cavity coupling strength $g$. In contrast, the SQUIDs in devices 2 and 3 are more symmetric, with the qubit frequency oscillating in such a way that it intersects with the cavity frequency around 5.5 GHz and 6 GHz, as shown in in Fig. \ref{FigS7} (b) and (c), respectively. Our flux-modulated cavity frequency data, presented in Fig. \ref{Fig6}, supports the simulations shown in Fig. \ref{FigS7}. Fig. \ref{Fig6} (a) shows the flux modulation of cavity frequency for all three devices, while Fig. \ref{Fig6} (b) provides the corresponding linecuts along the dashed lines in Fig. \ref{Fig6} (a). Since $f_{q,2}$ intersects the cavity frequency in a higher position in the modulation than $f_{q,3}$, as shown in Fig. \ref{FigS7} (b) and (c), the ratio between the flux regions where $f_{q}$ $<$  $f_{r}$ and $f_{q}$ $>$  $f_{r}$ is expected to differ between the two devices. This can be clearly observed in Fig. \ref{Fig6} (b), where the ratio between the lengths of line A (denoting the region where $f_{q}$ $<$  $f_{r}$) and line B (denoting the region where $f_{q}$ $>$  $f_{r}$) is larger in device 2 as compared to that in device 3. In contrast, $f_{q,1}$ oscillates well above the cavity frequency as shown in Fig. \ref{FigS7} (a), we expect the up-and-down modulation of $f_{q,1}$ will reflect on that in cavity frequency similarly. This is also observed in the leftmost panel of Fig. \ref{Fig6} (b), where the ratio between the length of line A and B is close to unity.

The estimated $g/2\pi$ is 41.76 MHz for device 2 and 68.5 MHz for device 3, indicating that energy exchange ($g/\pi$) between the qubit and the cavity would occur with periods of 11.97 ns for device 2 and 7.3 ns for device 3. Since $f_{q}$ intersects with $f_{r}$ in device 2 and 3, the fact that the avoided crossing was not observed in Fig. \ref{Fig3} (b) and (c) suggests that the qubit loses its coherence before a full round-trip exchange of energy can occur between the qubit and the cavity (see discussions in the final part in Appendix D). This suggests a qubit decay rate $\gamma$ is greater than the qubit-resonator energy exchange rate $g/\pi$, from which we can estimate the upper bound of $T_1$ = $1/\gamma$ $<$ 11.97 ns for device 2 and $T_1$ = $1/\gamma$ $<$ 7.3 ns for device 3. We noted a recent study demonstrating a gate-compatible 3D cavity architecture \cite{Xia2024,Kong2015}. In future efforts, it would be intriguing to incorporate gate-tunability into our SQUID design, allowing for electrical control to tune the maximum $f_{q}$ close to the cavity frequency. This could facilitate operation near the sweet spots within the strong dispersive regime, enabling state-dependent readout while suppressing flux noises. Furthermore, we anticipate by reducing the residue of conducting 2D materials on the substrate surface, replacing Si$/$SiO$_2$ substrates with Si or sapphire substrates to reduce the dielectric loss, and incorporating magnetic shielding along with infrared radiation filters could further improve the decoherence properties of our systems \cite{Oliver2013,OConnell112008,Corcoles2011,Barends2011}.

\maketitle
\section{IV. Summary and prospect}

In summary, we have demonstrated 3D cavity-compatible cQED devices incorporating graphene SQUIDs, capable of both DC and microwave characterizations. In microwave measurements, we observed Fraunhofer modulation in a device with asymmetric SQUID, whereas no Fraunhofer pattern was found in other devices with relatively symmetric SQUIDs. We have provided a schematic illustrating the relationship between DC SQUID critical current, qubit frequency, and cavity dispersive shift under the influence of the Fraunhofer effect. Based on the DC critical current analysis, we extracted information about $f_{q}$, $g$ and SQUID symmetry, which was then correlated with the distinct behavior of flux-modulated cavity frequency in three devices. Our 3D cavity platform can extend to topological materials, where MBS play a role and result in nontrivial 4$\pi$-period supercurrent and qubit frequency \cite{Badiane2013,Sun2022}. Under this scheme \cite{Chiu2024}, one can first probe the Shapiro steps to reveal the existence of non-trivial 4$\pi$-periodic ABSs \cite{Wiedenmann2016, Bocquillon2017, Li2018a}. The same device can then be placed in a 3D cavity to probe the 4$\pi$ modulation of qubit frequency using time-domain spectroscopy, with the possibility to avoid quasiparticle poisoning \cite{Sun2022}. 

\maketitle
\section{V. Acknowledgments}
Kuei-Lin Chiu would like to thank the funding support from National Science and Technology Council (Grant No. NSTC 109-2112-M-110-005-MY3 and NSTC 112-2112-M-110-017). Chung-Ting Ke and Yi-Chen Tsai would like to thank the funding support from National Science and Technology Council (Grant No. NSTC 110-2628-M-001-007). Kuei-Lin Chiu and Yen-Hsiang Lin also acknowledge support from the Center for Quantum Science and Technology (CQST) within the framework of the Higher Education Sprout Project by the Ministry of Education (MOE) in Taiwan. 

Kuei-Lin Chiu would like to thank Valla Fatemi and Yueh-Nan Chen for the useful suggestions on the manuscript.

\maketitle
\section{VI. Author contributions}
K. L. Chiu conceived the project. Y. Chang fabricated the devices under the supervision of C. T. Ke. Y. H. Chen and A. J. Lasrado calibrated the 3D cavity with input from K. L. Chiu and Y. H. Lin. A. J. Lasrado performed the microwave measurements under the supervision of K. L. Chiu and with contributions from C. H. Lo, T. Y. Hsu and Y. C. Chen. Y. Chang performed the DC transport measurements under the supervision of C. T. Ke and with contributions from Y. C. Tsai. Samina and A. J. Lasrado performed the capacitance simulation under the supervision of Y. H. Lin. K. L. Chiu and C. T. Ke co-supervise the project. 

\maketitle
\section{VII. Competing financial interests}
The authors declare no competing financial interests.

 \widetext
 \clearpage
 \begin{center}
 	\textbf{\large Appendices}
 \end{center}

 \setcounter{section}{0}
 \setcounter{table}{0}
 \makeatletter
 \renewcommand{\thesection}{S\arabic{section}}
 \renewcommand{\thesection}{\Roman{section}}

 \section{APPENDIX A: Device fabrication}

\begin{figure}[h]
 	\includegraphics[scale=0.35]{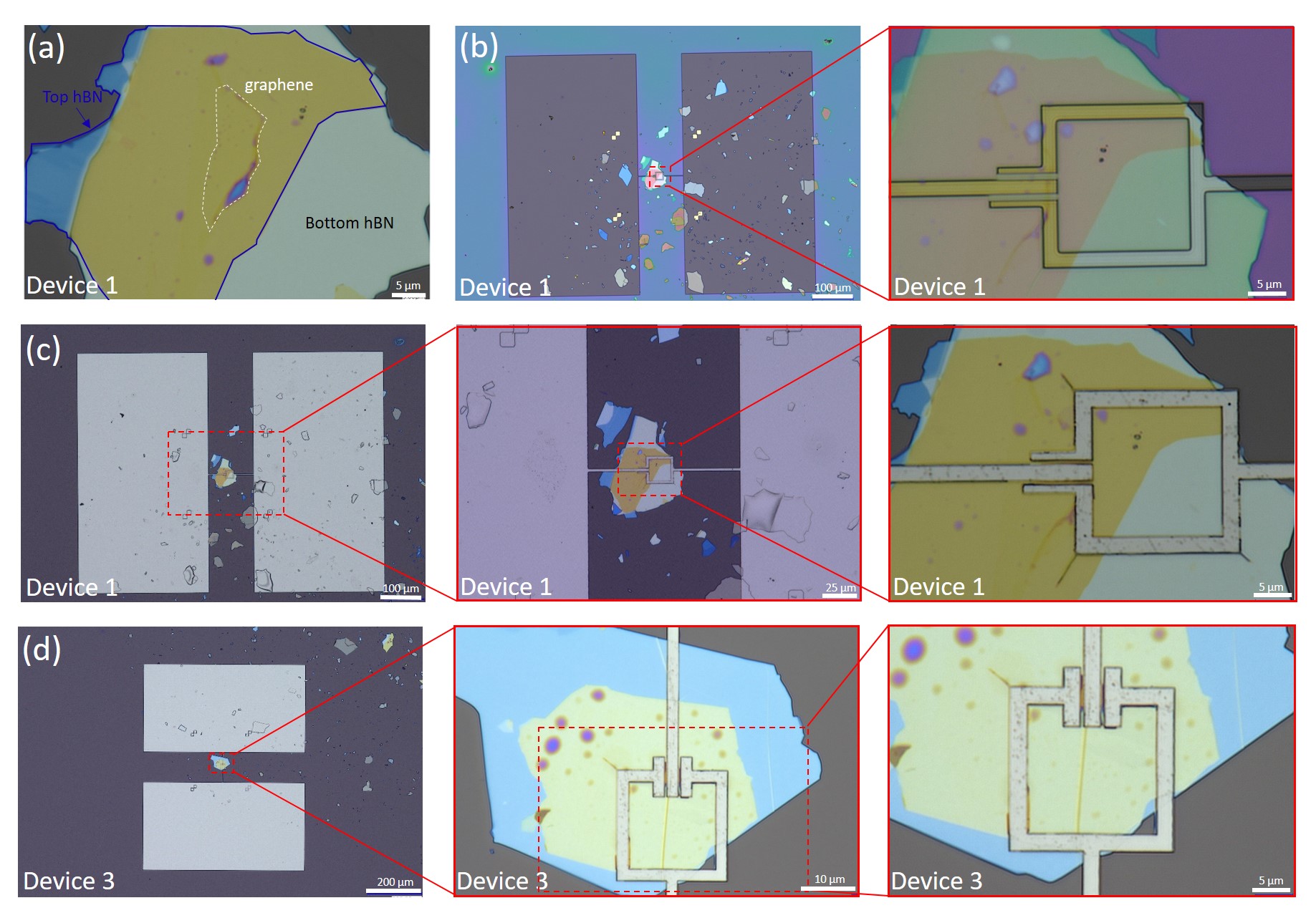}
 	\caption{Optical microscope images of device 1 at different stages during fabrication and the as-fabricated device 3. (a) Optical micrograph of hBN/Graphene/hBN sandwich structure after 2D material transfers. (b) Optical micrograph of device 1 after E-beam exposure and before ICP-RIE etching and Nb sputtering, with zoom-in image showing the SQUID structure. (c) Optical microscope images of the as-fabricated device 1 with different magnifications. (d) Optical microscope images of the as-fabricated device 3 with different magnifications.}      
 \label{FigS2}
 \end{figure}

We first prepare the hBN/graphene/hBN sandwich on intrinsic Si wafers capped with 90 nm SiO$_2$ (substrate) using the polymer-free dry transfer method \cite{R.2010,Mayorov2011,Haigh2012,Wang2013}. Fig. \ref{FigS2} (a) shows the optical micrograph of the as-transferred heterostructure (device 1), with a graphene flake encapsulated between two hBN layers. After spin coating of PMMA (A6, 500 rpm for 5 s then 4000 rpm for 55 s, bake at 170 $^\circ$C for 2 mins), electron beam lithography (EBL) was used to define the pattern for capacitor pads and SQUID contacts. After EBL exposure and developing, the optical micrograph of device 1 at this stage is shown in Fig. \ref{FigS2} (b). In our SQUIDs, we have designed three contacts across the graphene layer, with a contact width of 2 $\mu$m and a 500 nm gap between each other. The extension of SQUID contacts connects to a pair of capacitor pads, each with a dimension of 600 $\mu$m $\times$ 320 $\mu$m. In order to make edge contact to the encapsulated graphene, a Inductively Coupled Plasma Reactive-Ion Etching (ICP-RIE) technique was performed, using CHF$_3$ and O$_2$ gases with a ratio of 20:1 (power: 150 W and bias: 20 V), to selectively remove the hBN and graphene layers. This allows us to create the desired side contact geometry. Subsequently, 120 nm niobium (Nb) was sputtered (pressure: 3 mTorr, power: 25 W and sputtering rate: 6 nm/min) right after the etching process, ensuring minimal contact resistance. After lift-off, the optical micrograph of as-fabricated device 1 and device 3 are shown in Fig. \ref{FigS2} (c) and (d), respectively. 

 
\section{APPENDIX B: Measurement scheme}

\begin{figure}[!t]	
 	\includegraphics[scale=0.6]{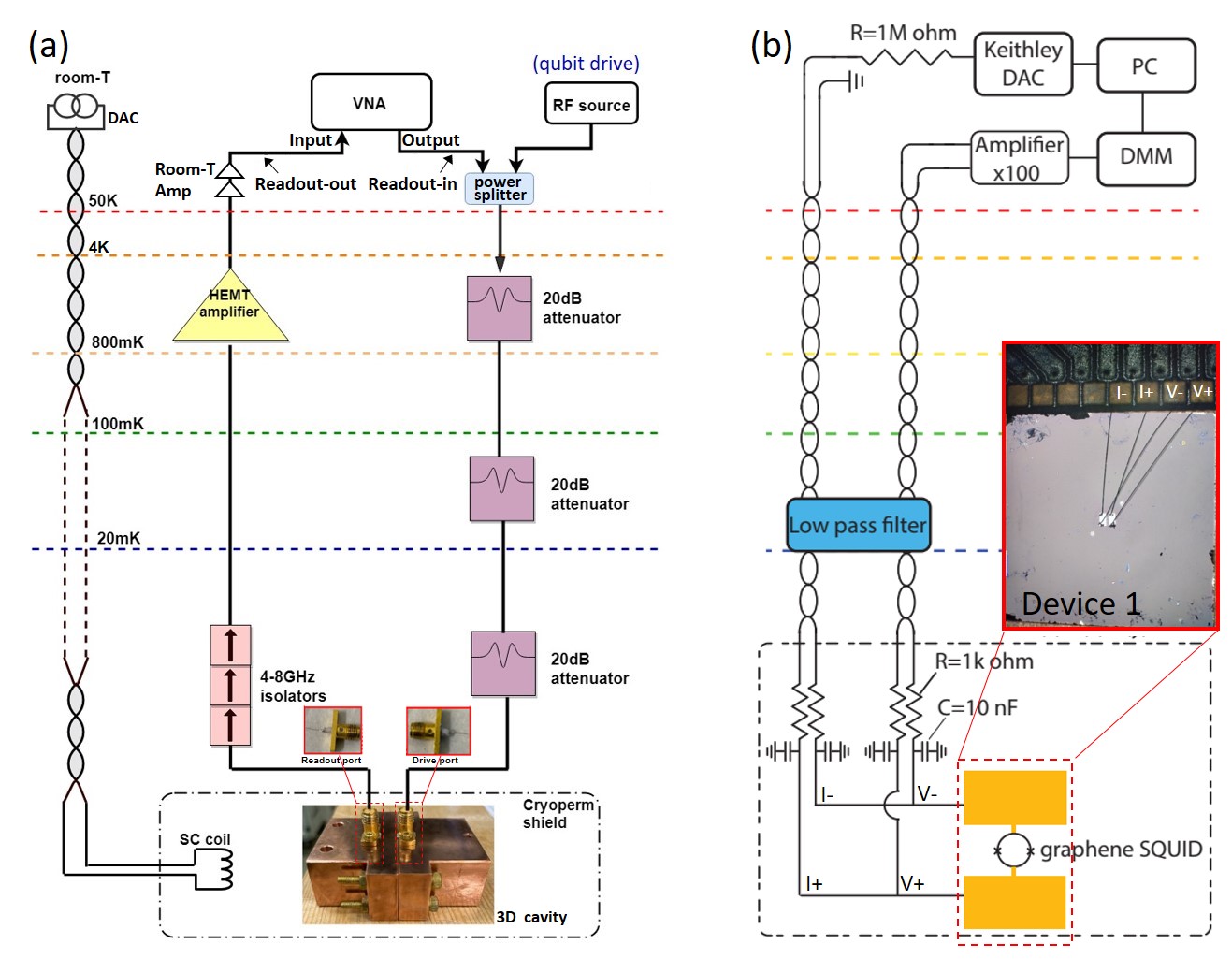}
 	\caption{(a) The measurement scheme used to perform the microwave characterizations, including power dependence, flux modulation and two-tone spectroscopy of our graphene-based superconducting quantum circuits. (b) The measurement scheme used to perform the DC transport measurements of the same devices that were previously characterized using the microwave measurement scheme shown in (a).}      
 	\label{FigS3}
 \end{figure}

Our 3D cavity devices allow both microwave measurements and DC transport measurements. All the experiments detailed in the main text were performed in a dilution refrigerator with a base temperature of 10 mK. In this section, we introduce our 3D cavities and describe the measurement schemes for both microwave [Fig. \ref{FigS3} (a)] and transport measurements [Fig. \ref{FigS3} (b)]. 

\subsection{1. 3D cavity design and calibration}

The 3D cavity is designed as a rectangular resonator, the resonance frequency of which is given by the formula: $f_{mnl}$ = $\frac{c}{2\pi\sqrt{\mu_r\epsilon_r}}$$\sqrt{(\frac{m\pi}{a})^2+(\frac{n\pi}{b})^2+(\frac{l\pi}{d})^2}$, where $a$, $b$ and $d$ represent the three-dimensional lengths of the rectangular resonator while $m$, $n$ and $l$ represent the mode numbers. Generally, the transmon placed in the center of the cavity chamber [Fig. \ref{Fig1} (a)] is primarily coupled with the electric field of TE$_{101}$ mode \cite{Nguyen2020}. The 3D copper cavity is milled from two pieces, one half consisting of a pin to drive qubits (drive port), and another with a pin to read qubits (readout port). The insertion depth of the pin within the ports determine pin’s quality factors. Typically the readout port with lower quality factor consists of a longer pin, which has higher visibility to the cavity photons, and thus couples strongly with the EM fields. In contrast, the drive port with higher quality factor consists of a shorter pin, thus limiting the leakage of photons through it. The quality factor of each pin is found by the formula of reflection parameter $S_{11} (f)$ = $\frac{2i(f-f_0)-f_0/Q_{ext}+f_0/Q_{int}}{2i(f-f_0)+f_0/Q_{ext}+f_0/Q_{int}}$, where $f_0$ is the cavity resonance frequency, $Q_{ext}$ is the Q value of the tested pin, and $Q_{int}$ is the internal Q value of the 3D cavity \cite{Nguyen2020}. We measure $S_{11}$ of each pin with Vector Network Analyzer (VNA). By applying the above formula, we fit the amplitude and phase curve to the measured data to determine $Q_{int}$ and $Q_{ext}$ of each pin. By carefully trimming pins and measuring the $Q_{ext}$, we generally reach a desired $Q_{ext}$ ratio of drive pin to the readout pin of about 3:1 \cite{Nguyen2020}, with specific values of around 5000:1500. For empty cavities, $Q_{int}$ is usually between 3000 and 3500 at room temperature and reaches around 20000 at 10 mK.

\subsection{2. Microwave measurements}
The measurement of microwave through the two-ports 3D copper cavity is performed with a transmission setup, in which two sets of coaxial lines are utilized: readout-in line serving as the input line while readout-out line as the output line, as shown in Fig. \ref{FigS3} (a). Readout-in line sends microwave from the output port of VNA (Keysight E5071C) to the drive port of 3D cavity. Subsequently, the transmitted signal from the readout port of 3D cavity travels through the readout-out line, finally reaching the input port of the VNA. By using a power splitter, readout-in line can also receive microwave signals from the RF source (Rohde Schwarz SGS100A) to drive the qubit for two-tone measurements. Since the entry of thermal photons from room temperature sources into the cryogenic environment can critically excite the qubits, the readout-in line is heavily attenuated with three 20 dB attenuators connected at different stages of dilution refrigerator. On the other hand, the readout-out line, which has no attenuation (0 dB), is connected to the High Electron Mobility Transistor (HEMT) amplifier at the 4K stage. The HEMT (LNF-LNC4 8C) has about 40 dB of gain and a noise temperature of 2K. To shield the device from thermal radiation from the HEMT amplifier, a set of three isolators are installed. The readout signal is further amplified by two room-temperature amplifiers (in total 40 dB gain) before going into the input of the VNA. In order to provide the necessary magnetic flux, a home-made superconducting coil is used, which is powered by an external DC source (digital-to-analog converter, DAC). The entire measurement setup is illustrated in Fig. \ref{FigS3} (a).

\subsection{3. DC transport measurements}
DC measurement is conducted within a dilution refrigerator operating at a base temperature of 10 mK, as illustrated in Fig. \ref{FigS3} (b). To facilitate precise four-wire measurements, the samples are affixed using four-wire connections, comprising a DC current source, two voltage probes, and a ground. The DC current source is established by combining a Keithley DAC voltage source with a range of $\pm$10 V and a room temperature 1 M$\Omega$ resistor. This setup allowed to generate a stable current in a range of $\pm$10 $\mu$A. Two voltage probes are positioned on either side of the SQUID device, and the resulting voltage difference was then amplified by a room-temperature voltage amplifier, featuring a fixed gain of 100. Throughout the manuscript, it is important to note that the current range and amplification factor remained consistent for all samples. To minimize the impact of external noise, twisted paired wiring is utilized from room temperature down to the mixing chamber level. To further enhance signal quality, two stages of low-pass RC filters are deployed at room temperature and on the printed circuit board (PCB). Additionally, a $\pi$ filter from QDevil is incorporated at the mixing chamber level within the dilution refrigerator, serving the dual purpose of thermalization and noise reduction. All measurements and data acquisition are orchestrated through a dedicated PC.

\section{APPENDIX C: SQUID oscillations with Fraunhofer effect}

\subsection{1. Fraunhofer pattern of symmetric SQUIDs}
In a symmetric SQUID whose interference pattern is modulated by the Fraunhofer diffraction patterns of each JJ, the total critical current can be described as \cite{Qu2012}:

\begin{eqnarray}
I_{c}(B) &=& 2I_{c}(0) \left|sin(\frac{\pi\Phi_J}{\phi_0})/(\frac{\pi\Phi_J}{\phi_0})\right|\left|cos\frac{\pi\Phi}{\phi_0}\right| \label{eqn 1}
\end{eqnarray}
where $I_{c}(0)$ is the critical current of each junction at zero magnetic field, $\Phi_J$ is the flux threading
through the single junction area, $\Phi$ is the flux threading through the loop area of SQUID and $\phi_0$ = $h/2e$ is the flux quanta. The first term in equation (1) represents the Fraunhofer pattern which consists of a central lobe and a series of sub-lobes, while the second term represents SQUID oscillations. The central lobe in Fraunhofer pattern contains twice the number of SQUID oscillations in other sub-lobes. In equation (1), the ratio between $\Phi_J$ and $\Phi$, meaning the ratio between JJ's area and SQUID loop area, determines how many SQUID oscillations reside in a Fraunhofer lobe. In Fig. \ref{FigS5} (a) and (b), we plot the function $y = \left|sin(x)/x\right| \left|cos(A\ast x)\right|$, in which x represent the flux and the factor $A$, denoting the the area ratio between SQUID loop and JJ, is set to 10 and 40, respectively. This is to mimic the data as shown in Fig. \ref{Fig3} (a) and (d), in which 10 and 41 SQUID oscillations reside in a Fraunhofer sub-lobe, respectively. Note that in a real case, the period of SQUID oscillation is fixed (as the designed SQUID loop area for three devices are the same), while the period of JJ's Fraunhofer oscillation varies due to the differences in JJ's area (i.e., width of graphene flake) from device to device. Here, we keep the period of the JJ's Fraunhofer oscillation constant while varying the period of the SQUID oscillation to facilitate a straightforward comparison of devices with different SQUID loop-to-junction area ratios.

\subsection{2. Fraunhofer pattern of asymmetric SQUIDs}
For an asymmetric SQUID with Fraunhofer pattern modulated by junctions of two different sizes (the case of device 1), we construct the function as $y = \left|sin(x)/x\right| \left|sin(4x)/x\right| \left|cos(40\ast x)\right|$. Here, the first term represents the Fraunhofer modulation from the smaller junction (JJ$_1$), the second term accounts for the modulation from the larger junction (JJ$_2$), and the third term describes the SQUID oscillations. Note that the function form of SQUID oscillation is still based on a symmetric SQUID, but it does not affect our analysis below, as our primary focus is on discussing the interplay between the two Fraunhofer patterns. The ratio between the factors 1:4:40 denotes the area ratio between JJ$_1$, JJ$_2$ and SQUID loop, which is based on our analysis ($I_{C1}$ : $I_{C2}$ = 1:4 for device 1) in Appendix E. The function is plotted in Fig. \ref{FigS5} (c), with red curve indicating the total function value and blue (green) curve indicating the Fraunhofer pattern of JJ$_1$ (JJ$_2$). As can be seen, the SQUID oscillations, which are modulated by the short-period Fraunhofer pattern from JJ$_2$, are further modulated by the long-period Fraunhofer pattern from JJ$_1$. In our microwave measurements, as shown in Fig. \ref{Fig3} (a), we observed the Fraunhofer modulation of JJ$_2$ with sub-lobes containing 10 SQUID oscillations. However, we did not observe the full behavior depicted in Fig. \ref{FigS5} (c) in our transport data, shown in Fig. \ref{Fig3} (d). We attribute this to a possible offset between the two Fraunhofer patterns of JJ$_1$ and JJ$_2$. As an example, if JJ$_1$'s Fraunhofer central lobe is shifted to JJ$_2$'s Fraunhofer sub-lobes, one can imagine the SQUID oscillations in JJ$_1$'s central lobe will not be much modulated by JJ$_2$'s sub-lobes, as the modulation in the sub-lobe is much weaker. In this case, we observe only a dominant Fraunhofer pattern without the additional Fraunhofer modulation from the other junction. We found relevant clues in the third cool down of device 1, in which transport measurement was performed again as shown in Fig. \ref{FigS4}. In Fig. \ref{FigS4} (a), the main features of Fraunhofer pattern remain almost the same compared to Fig. \ref{Fig3} (d), with now 73 SQUID oscillations in the central lobe instead of 82 observed in the second cool down. However, this time we push the magnetic field range far away from the central lobe region, as shown in Fig. \ref{FigS4} (b). As observed, there is a new period of Fraunhofer lobe containing only 22 SQUID oscillations, which is clearly not half of 73 and cannot be explained by assuming a symmetric SQUID. This indicates the presence of another junction with a larger area compared to the one responsible for the 73 SQUID oscillations in the Fraunhofer central lobe.

\begin{figure}[!t]	
\includegraphics[scale=0.55]{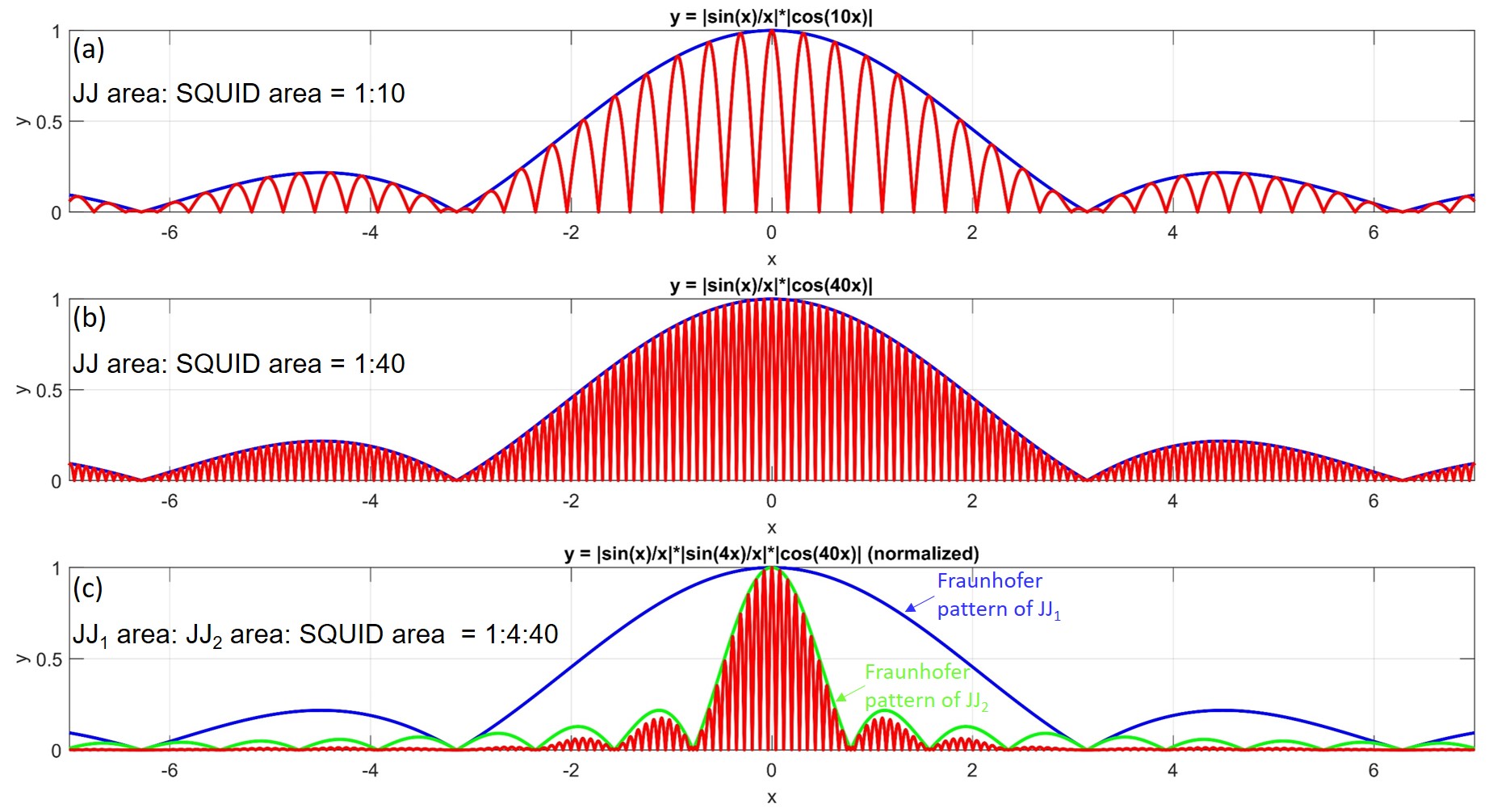}
\caption{(a) Schematic simulation of SQUID critical current (red curve) modulated by flux under the influence of Fraunhofer effect (blue curve), with 20 SQUID oscillations in the Fraunhofer central lobe. (b) The same as (a) but with 80 oscillations in the Fraunhofer central lobe. (c) Critical current (red curve) modulated by flux from an asymmetric SQUID, influenced by the Fraunhofer effect from two JJs with a area ratio of 1:4. The blue curve denotes the Fraunhofer pattern of JJ$_1$, while the green curve denotes that of JJ$_2$. All curves are normalized to 1.}      
\label{FigS5}
\end{figure}

\subsection{3. Dispersive shift modulated by Fraunhofer pattern}
In Fig. \ref{Fig4}(b), we schematically simulate the dispersive shift $\chi = -g^2/2\pi(f_q - f_{r})$ under the influence of Fraunhofer effect. We aim to capture how Fraunhofer pattern modifies the periodically flux-modulated cavity frequency in SQUID-based transmons. To this purpose and for simplicity, we set $g^{2}/2\pi$ = 1, and used $f_{r}$ = 0 as a reference point. In Fig. \ref{FigS7}, we have shown the oscillation of $f_{q}$ with respect to the cavity frequency for all devices. Our primary interest lies in the Fraunhofer modulation of the maximum $f_{q}$, which is well above $f_{r}$. Therefore, we set $f_{r}$ = 0 as a reference and focus solely on the regime where $f_{q}$ $>$  $f_{r}$ in Fig. \ref{Fig4}(b). We disregard the function values at specific points where $f_{q}$ $\approx$ 0, which cause divergence in the function values. We understand that the symmetry of SQUID for device 1-3 is different, which results in different critical current minimum and determines whether $f_{q}$ intersects with cavity frequency (see Fig. \ref{FigS7}). However, since we are only interested in the Fraunhofer modulation on the maximal $f_{q}$, the results from Fig. \ref{Fig4}(b) are still valid regardless of SQUID symmetry.

\begin{figure}[!t]	
\includegraphics[scale=0.46]{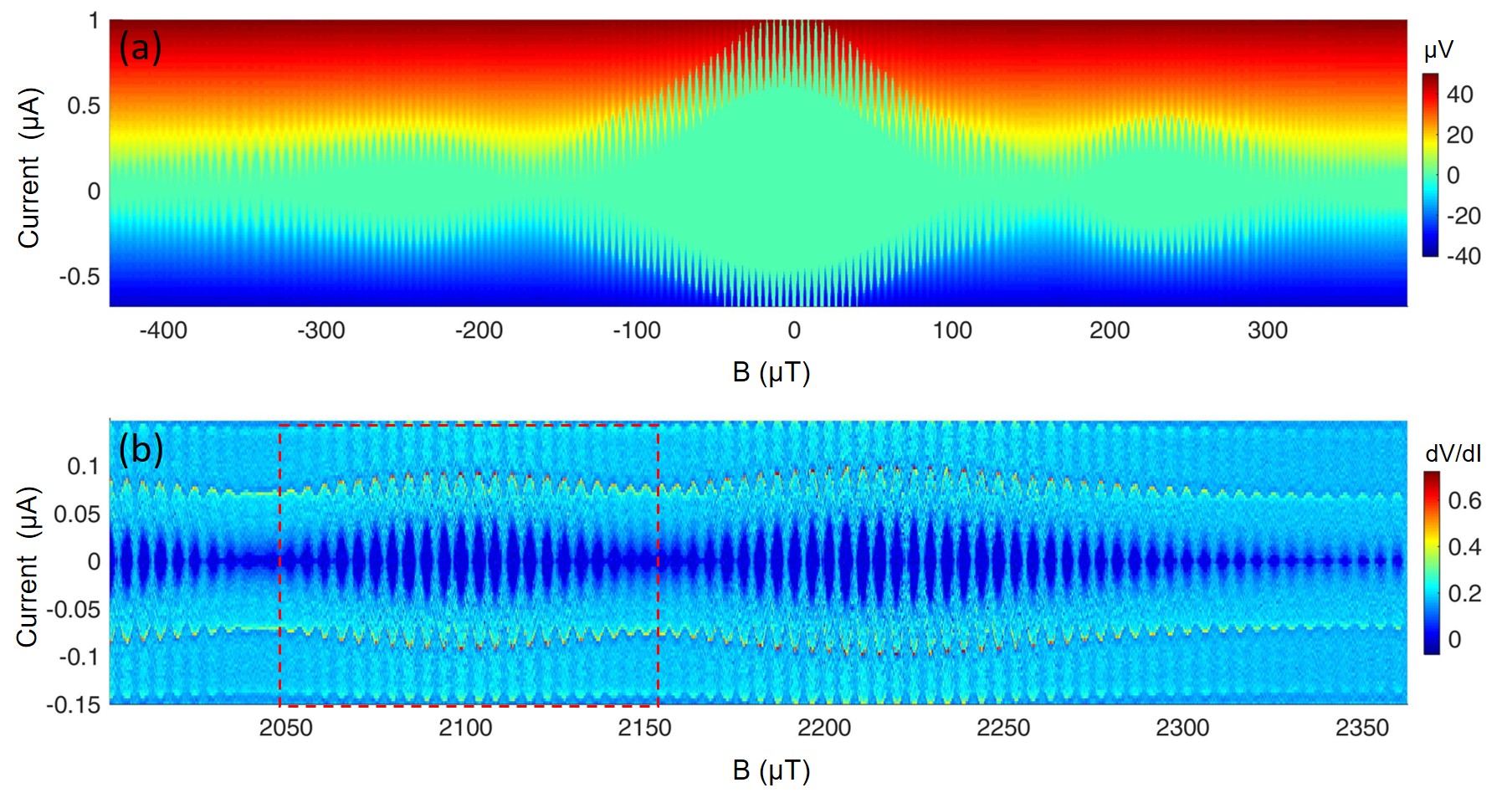}
\caption{(a) Flux modulation of SQUID critical current for device 1 in it's third cool-down. The colorscale represents the DC voltage across the junctions. (b) A Fraunhofer lobe (indicated by the red dashed square) far from the central lobe, containing 22 SQUID oscillations. The raw data is differentiated along y-axis (current direction) in order to resolve the oscillations better.}      
\label{FigS4}
\end{figure}


\section{APPENDIX D: Estimation of qubit parameters}

\begin{figure}[!t]	
\includegraphics[scale=0.58]{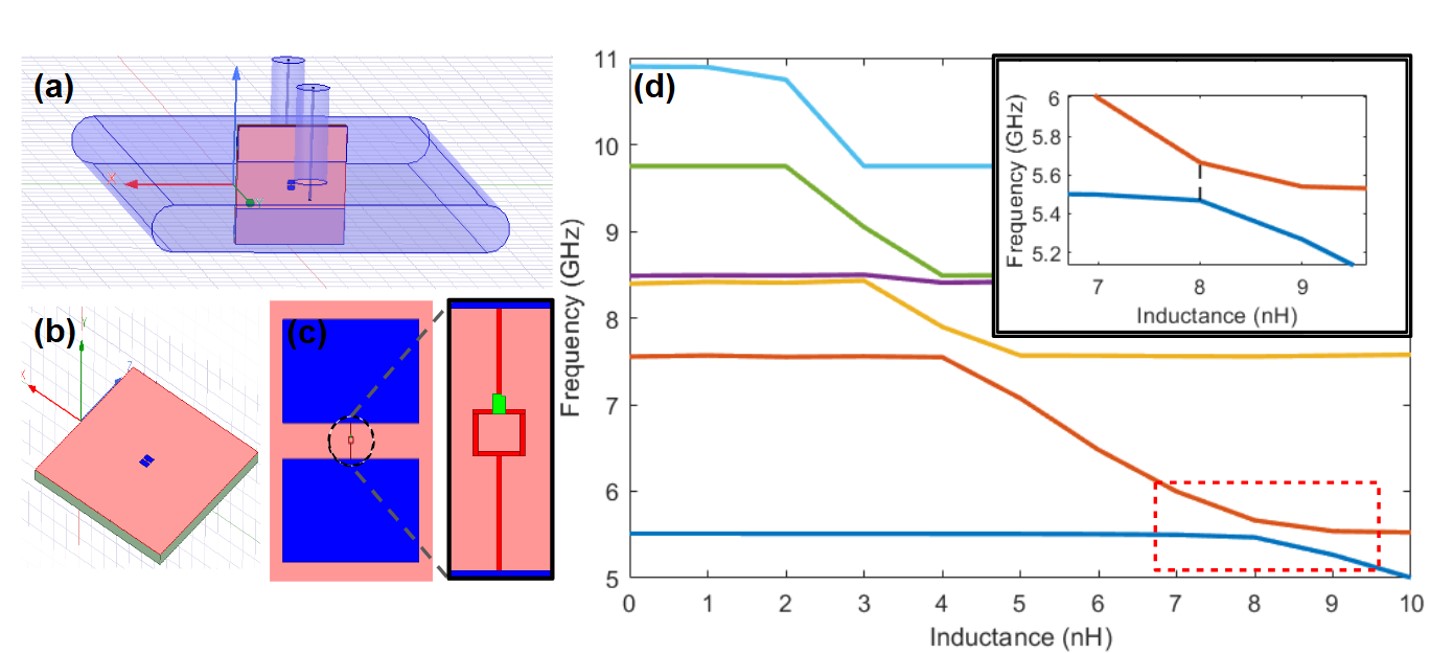}
\caption{Simulation of qubit-cavity coupling strength $g$. (a) Transmon chip placed inside the 3D cavity. (b) Substrate containing transmon qubit at its center. (c) Transmon with a pair of capacitor pads (blue boxes) and a SQUID (red) positioned between them. Inset shows the close-up view of the SQUID, with the JJ (green box).  (d) Eigen-mode simulation of first six modes, with the inset showing the avoided crossing between the first two modes, which determines cavity-qubit coupling strength $g$.}      
\label{FigS9}
\end{figure}

\subsection{1. Estimation of $E_C$, $E_J$, and $g$}
By using Ansys Maxwell 3D software, we simulated the intrinsic capacitance between the capacitor pads along with non-connected SQUID contacts, including the effects of the cavity. By applying the voltage of 1 V for one pad while keeping the other pad at 0 V, the capacitance between the pads is determined to be 92 fF. We further simulate the qubit-cavity coupling strength $g$ by using Ansys HFSS. Fig. \ref{FigS9} (a) features a transmon chip placed inside a 5.8 GHz cavity, which is enclosed by a perfect conductor boundary. Fig. \ref{FigS9} (b) depicts a transmon situated on the Si/SiO$_2$ substrate while Fig. \ref{FigS9} (c) shows the transmon equipped with two large capacitor pads (blue boxes), with a SQUID loop bridging between them. The JJ is specially modeled by assigning the boundary of a conducting box as lumped RLC with parallel inductance. This simulates the effect of JJ. In this design, all the metallic parts are made up of perfect conductor material. For simulating the cavity-qubit coupling strength $g$, we modulate the qubit frequency by changing the value of junction inductance. In a coupled system, as the qubit frequency approaches the cavity frequency, there will be deviation in cavity frequency resulting in an avoided crossing of qubit and cavity resonant frequencies. We have computed the first six eigen-mode solutions (Fig. \ref{FigS9} (d)), with the first mode (blue curve) corresponding to cavity frequency, and the second mode (red curve) corresponding to the qubit frequency. Inset shows the avoided crossing (black dashed line) between the first two modes that is highlighted in the red dashed frame. From the data, the qubit-cavity coupling strength is simulated as $g$ = 98 MHz at junction inductance around 8 nH.
 
Based on the dispersive shift obtained from the power dependence measurements, along with the SQUID critical current obtained from DC transport, we can also estimate the qubit-cavity coupling strength $g$ based on experimental data using $\Delta/2\pi = f_q - f_r$ and $\chi = g^2/\Delta$. Since qubit frequency $f_q$ $\approx$ $\sqrt{8 E_J E_C}/h$, we need to know both charging energy $E_C$ = $\frac{e^2}{2C}$ and Josephson energy $E_J$ = $\frac{\Phi_0I_C}{2\pi}$. From our simulation, we obtained a capacitance $C$ = 92 fF, which allows us to estimate $E_C/h$ $\approx$ 210.5 MHz. Next, we need to know SQUID critical current $I_C$ to estimate $E_J$. From Fig. \ref{Fig3} (a), we observe that three lobes contain an equal number of SQUID oscillations, suggesting that they originate from Fraunhofer sub-lobes rather than the central lobe. This could be due to the remanent magnetic field in our homemade superconducting coil, which causes a shift in the probing regime. Similarly, the flux modulation observed in Fig. \ref{Fig3} (b) and Fig. \ref{Fig3} (c) also originates from the Fraunhofer sub-lobes. These sub-lobes have a smaller impact on SQUID oscillations compared to the central lobe, as evident in the transport data shown in Fig. \ref{Fig3} (e) and Fig. \ref{Fig3} (f). Thus, we choose the critical current in the first Fraunhofer sub-lobe in the transport data as the upper bound of $I_C$, i.e., $I_C$ $\approx$ 0.2 $\mu$A for device 1 [Fig. \ref{Fig3}(d)], $I_C$ $\approx$ 0.1 $\mu$A for device 2 [Fig. \ref{Fig3}(e)] and $I_C$ $\approx$ 0.2 $\mu$A for device 3 [Fig. \ref{Fig3}(f)]. Hence, $E_J/h$ $\approx$ 99.337 GHz for device 1, $E_J/h$ $\approx$ 49.668 GHz for device 2 and $E_J/h$ $\approx$ 99.337 GHz for device 3. Combined all the parameters, we can estimate qubit frequency $f_q$ $\approx$ 12.935 GHz and $\Delta/2\pi = f_q - f_r$ = 12.935 GHz - 5.4955 GHz $\approx$ 7.44 GHz for device 1, $f_q$ $\approx$ 9.146 GHz and $\Delta/2\pi = f_q - f_r$ = 9.146 GHz $-$ 5.5135 GHz $\approx$ 3.63 GHz for device 2 and $f_q$ $\approx$ 12.935 GHz and $\Delta/2\pi = f_q - f_r$ = 12.935 GHz $-$ 6.034 GHz $\approx$ 6.9 GHz for device 3. Making $2\pi$$\times$13.67 MHz = $\frac{g^2}{2\pi\times7.44 GHz}$, $g/2\pi$ = 318.9 MHz can be inferred for device 1; $2\pi$$\times$0.48 MHz = $\frac{g^2}{2\pi\times3.63 GHz}$, $g/2\pi$ = 41.76 MHz can be inferred for device 2; $2\pi$$\times$0.68 MHz = $\frac{g^2}{2\pi\times6.9 GHz}$, $g/2\pi$ = 68.5 MHz can be inferred for device 3. Note that all three devices share the same capacitor and SQUID contact designs, which should result in similar coupling strength $g$, assuming the capacitor structure is placed at the center of the cavity in each case. The exact cause of the variation in $g$ is not entirely clear. We speculate that differences in the distribution of residual metal and graphite flakes after the fabrication process might play a role. Because we placed device 1 and device 2 in the same cavity (5.5 GHz) with the pins untouched (i.e., the same external Q), but the resulting total Q is 390 for device 1 while 835 for device 2, obtained from the 0 dBm linecut in Fig. \ref{Fig1}(f) and 0 dBm linecut in the top panel of Fig. \ref{Fig5}(b). We can extract an internal Q $\approx$ 612 for device 1 and Q $\approx$ 3682 for device 2, using the identical external Q values for the pins (4707.5 : 1401.6). From the photo of device 1 [Fig. \ref{FigS2}(c)] and device 2 [Fig. \ref{Fig1}(b)], one can clearly see that device 1 has much more 2D material residues compared to device 2. We have also measured 3D cavities loaded with Si/SiO$_2$ substrates containing exfoliated hBN and exfoliated graphite for comparison (data not shown). We found that substrates with exfoliated hBN have minimal impact on the internal Q compared to bare Si/SiO$_2$ substrates, whereas substrates with exfoliated graphite significantly reduce the internal Q by two orders of magnitude. Based on these observations, we speculate that the remaining metallic graphite pieces interact with microwave, thus amend the distribution of electromagnetic field inside the cavity and enhance the local electric field, causing the variation in qubit-cavity coupling strength observed across different devices.

\subsection{2. The absence of Rabi splitting}
In the next section, we found that the qubit frequency in device 2 and device 3 intersects with the cavity frequency [Fig. \ref{FigS7} (b) and (c)], but Rabi splitting was not observed in the flux-tuning data shown in Fig. \ref{Fig3}(b) and (c). This indicates that device 2 and device 3 are not in the strong coupling regime: $g$ $\gg$ $\kappa$ and $\gamma$, where $\kappa$ is the cavity decay rate and $\gamma$ is the qubit total decay rate \cite{Blais2004}. From the power dependence measurements shown in the top panels of Fig. \ref{Fig5}(b) and (c), we can estimate $\kappa/2\pi$ $\approx$ 10 MHz for device 2 and $\kappa/2\pi$ $\approx$ 20 MHz for device 3, based on the full width at half maximum (FWHM) of the cavity response at low powers. Thus, in both devices, $g/2\pi$ is large enough compared to $\kappa/2\pi$ (41.76 MHz $>$ 10 MHz for device 2 and 68.5 MHz $>$ 20 MHz for device 3), while Rabi splitting is still absent. This suggest that $g$ is not greater than $\gamma$, from which we can estimate the upper bound of qubit coherence time as discussed in the main text.

\section{APPENDIX E: SQUID symmetry analysis}

\begin{figure}[!t]	
\includegraphics[scale=0.42]{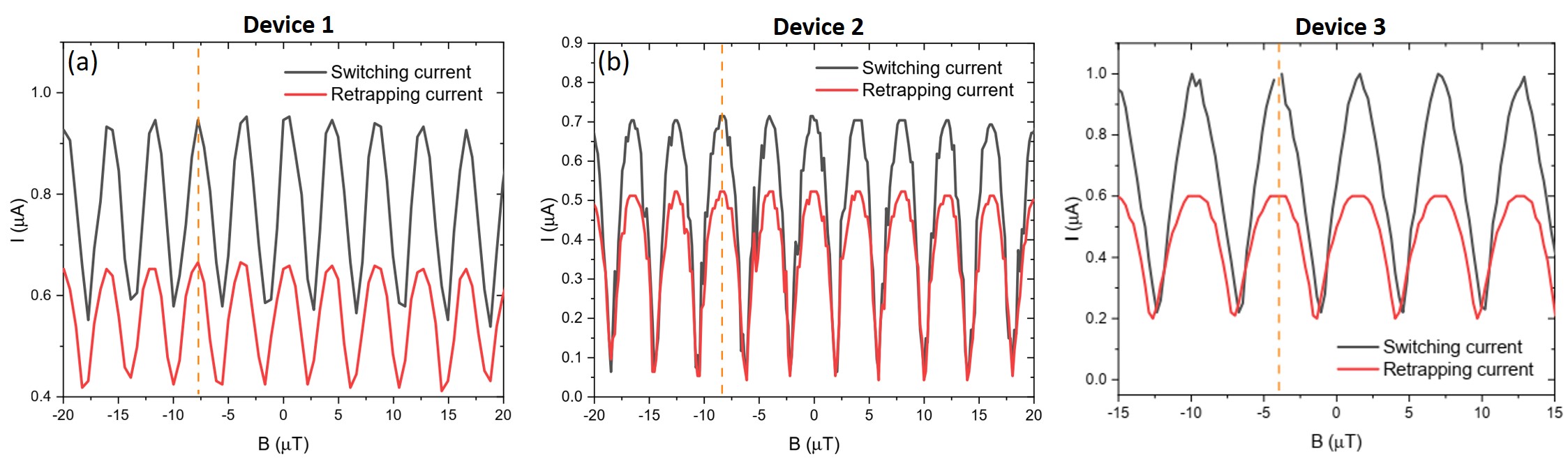}
\caption{The switching current and flipped retrapping current for (a) device 1, (b) device 2 and (c) device 3 . The data is extracted from the data shown in Fig. \ref{Fig3} (d), (e) and (f).}      
\label{FigS6}
\end{figure}

To give a qualitative analysis of the supercurrent in the individual JJ of the SQUID, we plot the switching current and retrapping current as the function of the magnetic field as shown in Fig. \ref{FigS6} for device 1, 2 and 3. Based on the result of the current-to-phase (CPR) relationship in Fig. \ref{FigS6}, we suggest that the inductance effect is small enough to allow for an analysis of the supercurrent in each JJ. Firstly, the supercurrent shows a symmetric shape with respect to phase (magnetic field) without skewness for all devices (see bottom panels of Fig. \ref{Fig5}). The asymmetry of CPR can come from the effect of the screening parameter $\beta$, which can be written as $\beta$ = $I_{max}\cdot L/\phi_0$, where $I_{max}$ is the maximum current through SQUID, $L$ is the inductance, and $\phi_0$ is the magnetic flux quanta. With a symmetric CPR (i.e, small $\beta$), this suggests that the influence of the inductance in the SQUID is negligible. Secondly, we compare the phase offset between switching and retrapping current, defined as $\Delta\phi = 2L(I_1-I_2)$, where $I_1$ and $I_2$ are supercurrents for two JJs in the SQUID. In Fig. \ref{FigS6}, we plot the switching and retrapping current for all three devices by flipping the retrapping current to the positive side. From the orange dashed lines, we see almost no phase offset from switching respect with retrapping current in all devices. The small phase difference indicates that they are in either symmetric supercurrent case $I_1$ = $I_2$ or small loop inductance $L$ case \cite{Nanda2017}. Clearly, for all devices, they are not in the case of $I_1$ = $I_2$, as the critical current does not reach zero (especially for device 1) as shown in the bottom panels of Fig. \ref{Fig5}, from which we know $L$ is close to zero. Thus, based on the two analyses, we conclude that the inductance is relatively small in our devices. This allows us to provide a rough estimation of the supercurrent for SQUID's JJs without complex calculation of the inductance of the SQUID system.

\begin{figure}[!t]	
\includegraphics[scale=0.55]{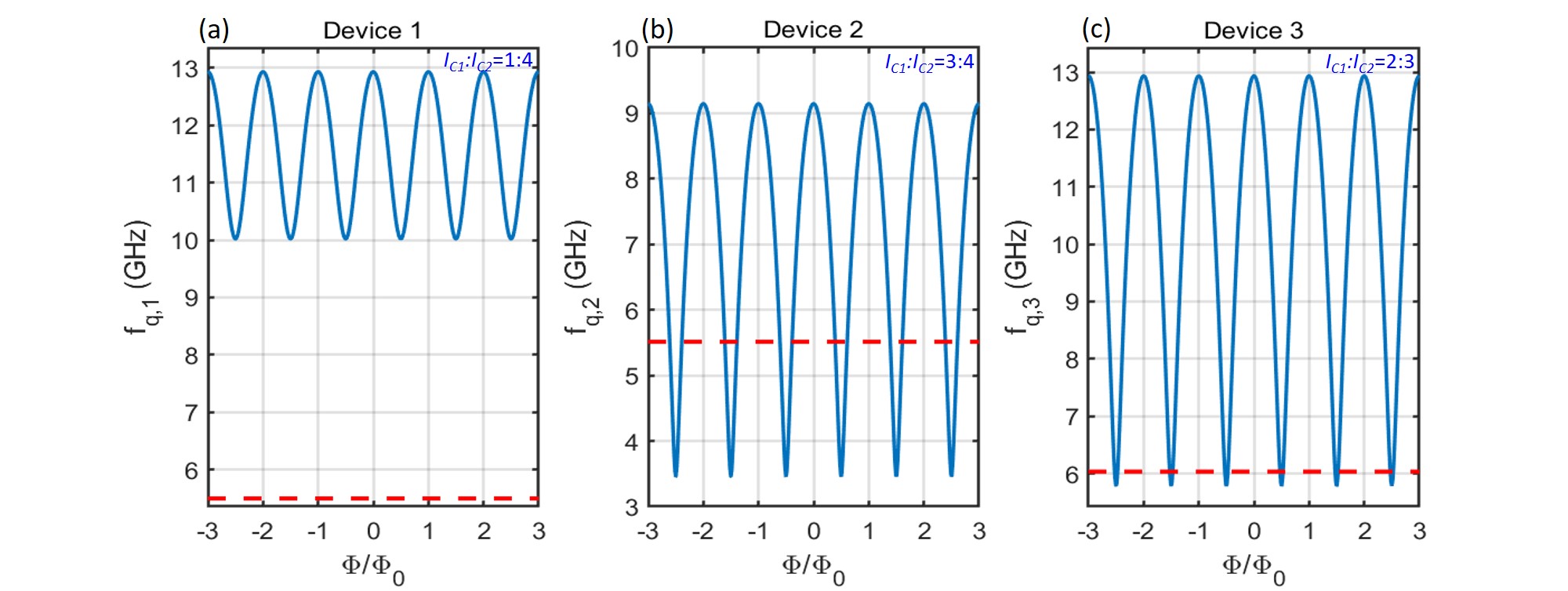}
\caption{Simulated qubit frequency modulation with respect to applied flux for (a) device 1, (b) device 2 and (c) device 3. The red dashed lines indicate the cavity frequencies.}      
\label{FigS7}
\end{figure}

We follow ref. \cite{Maier2015} to analyze the symmetry of SQUID in our devices. We chose the SQUID oscillations around $B$ = 0 T, as shown in the bottom panels of Fig. \ref{Fig5}, for analysis. For two junctions with different critical currents $I_{C,i}$, $i$ = 1, 2, the maximal SQUID critical current $I_{C,max}$ $=$ $I_{C,1} + I_{C,2}$, while the minimal SQUID critical current $I_{C,min}$ $=$ $I_{C,2} - I_{C,1}$. In device 1, $I_{C,max}$ $\approx$ 1 $\mu$A and $I_{C,min}$ $\approx$ 0.6 $\mu$A, from which we deduce $I_{C,1}$ $\approx$ 0.2 $\mu$A and $I_{C,2}$ $\approx$ 0.8 $\mu$A. Similarly, $I_{C,max}$ $\approx$ 0.7 $\mu$A (1 $\mu$A) and $I_{C,min}$ $\approx$ 0.1 $\mu$A (0.2 $\mu$A) for device 2 (3), from which we obtain $I_{C,1}$ $\approx$ 0.3 $\mu$A (0.4 $\mu$A) and $I_{C,2}$ $\approx$ 0.4 $\mu$A (0.6 $\mu$A) for device 2 (3). Thus, since $E_J$ $\propto$ $I_C$, $E_{J1}$:$E_{J2}$ is 1:4 for device 1, 3:4 for device 2 and 2:3 for device 3, respectively. Using $f_q$ $\approx$ $\sqrt{8 E_J E_C}/h$ and $E_J (\Phi)$ = $E_{J\Sigma}$ $\sqrt{cos^2(\Phi)+d^2sin^2(\Phi)}$ as depicted in the main text \cite{Krantz2020}, where $\gamma$ = 4 (4/3 and 3/2) and $d$ = 3/5 (1/7 and 1/5) for device 1 (2 and 3), we plot $f_q$ as a function of $\Phi$ in Fig. \ref{FigS7}. Note that we have combined all the non-$\Phi$-related parameters ($E_{J\Sigma}$$\sqrt{8 E_C}/h$) to match $f_{q,max}$ = 12.935 GHz, 9.146 GHz and 12.935 GHz for device 1, 2 and 3, as estimated in Appendix D. We did this as we are primarily interested in how SQUID symmetry (the ratio between $I_{C1}$ and $I_{C2}$) affects the oscillation depth of qubit frequency. As can be seen in Fig. \ref{FigS7}, $f_{q,1}$ oscillates weakly between 12.935 GHz and 10.02 GHz without intersecting with cavity frequency at 5.4955 GHz, while $f_{q,2}$ and $f_{q,3}$ from more symmetric SQUIDs oscillate more strongly and intersect with cavity frequencies at 5.5135 GHz and 6.034 GHz, respectively.

If we estimate the ratio between the length of A and B in Fig. 5, it is about 1:7.5 for device 2 and 1:17.83 for device 3. The same ratio estimated from Fig. \ref{FigS7} is 1:3.56 for device 2 and 1:16.88 for device 3. The experimentally estimated ratio is smaller (1:7.5) compared with the value based on simulation (1:3.56) for device 2, while the ratio is similar for device 3. This implies that our estimated $f_{q,2}$ in Fig. \ref{FigS7} should overall shift upwards and intersect with cavity frequency in a lower position of modulation. Since we estimated the qubit frequency based on critical current in transport measurements, there is a possibility that the actual critical current associated with qubit frequency may differ in microwave measurements, since it is in different cool down. However, this discrepancy does not affect our SQUID symmetry analysis and its correlation between DC and microwave measurements.

\section{APPENDIX F: Shapiro step measurements}

\begin{figure}[!t]	
\includegraphics[scale=0.42]{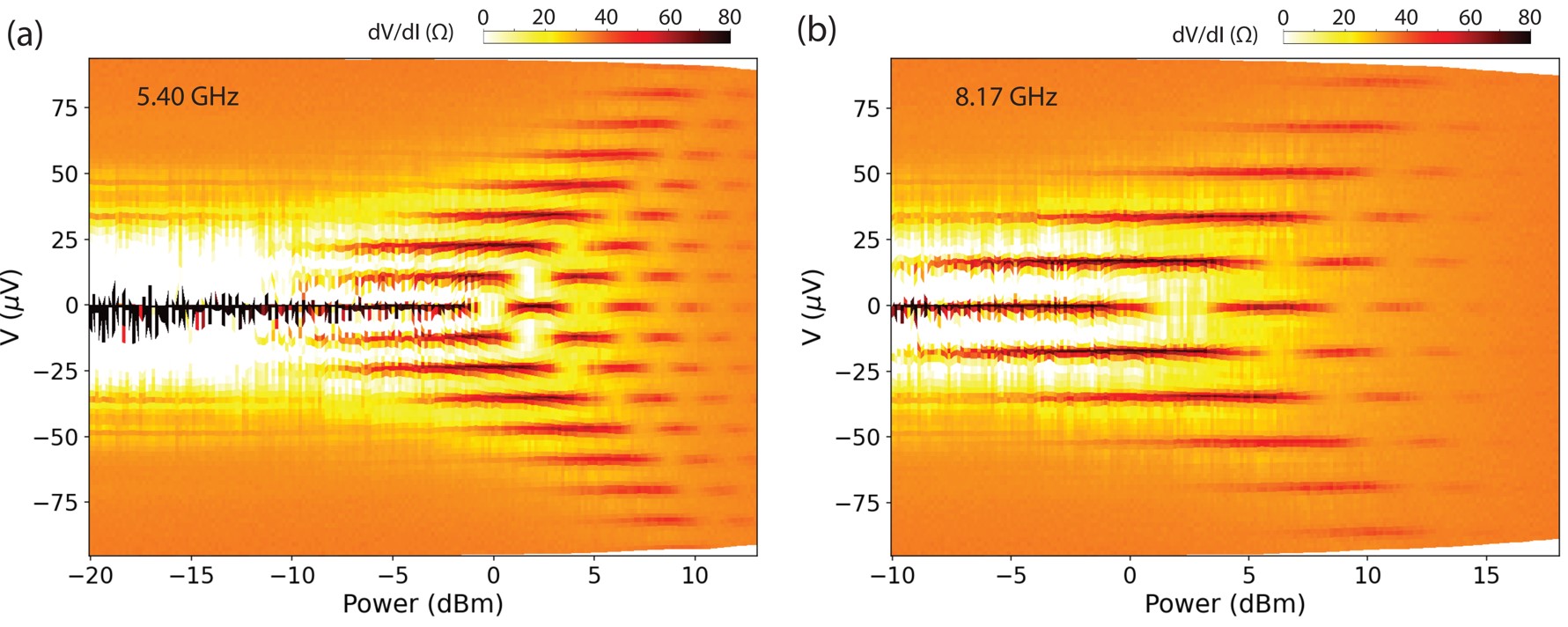}
\caption{Power dependence of Shapiro steps. We conduct Shapiro step measurements with two different frequencies, namely (a) 5.40 GHz and (b) 8.17 GHz. We can see that there are fewer Shapiro steps as the frequency increases. Increased power enhances the probability of tunneling processes resulting in more steps. However, increasing power beyond a certain point may not significantly improve the visibility of Shapiro's steps due to the heating effect of the samples.}      
\label{FigS8}
\end{figure}

As mentioned in the introduction of the main text, the absence of the n=1 Shapiro step indicates the presence of a nontrivial 4$\pi$-periodic supercurrent in topological Josephson junctions. In this section, we demonstrate that Shapiro step measurements can be implemented using our device architectures. We apply radio frequency (RF) signals onto our graphene-based JJs through a signal generator up to 20 GHz. The RF signal is attenuated at the 4K plate with a -20 dB attenuator then enter to the JJs $via$ an RF cable without further attenuation. By terminating the RF cable on top of our devices we are able to shine the FR photons onto the junctions. In Fig. \ref{FigS8} (a) and (b), we show the resistance maps of Shapiro steps for device 1 at RF frequencies of 5.40 GHz and 8.17 GHz, respectively. These two maps are derived from I-V measurements at different RF powers, from which we calculate the differential resistance based on the I-V curves. Finally, the map is replotted as the function of measured voltage versus RF power. A clear step feature at quantized voltage values can be observed. In device 1, the supercurrent is highly asymmetric as discussed in the main text. We measured Shapiro steps at a zero magnetic field (zero flux tuning), so the supercurrent is roughly at its maximum value of the graphene SQUID device. Due to the large asymmetry in device 1, the AC Josephson effect is presumably dominated by the single junction with a larger critical current. As a result, we primarily observed Shapiro steps from this single junction without detecting a more complex structure in device 1. These results demonstrate that AC Josephson effect can be realized in our 3D cavity-compatible superconducting quantum circuit devices.

\end{document}